\documentclass[a4paper,11pt]{article}
\usepackage{jheppub} 
\usepackage{amsmath,amssymb,color}
\usepackage{lineno}
\usepackage{booktabs}
\usepackage[T1]{fontenc}
\def\td{\tilde}

\newcommand{\be}{\begin{equation}}
\newcommand{\ee}{\end{equation}}
\newcommand{\bald} {\begin{aligned}}
\newcommand{\eald}{\end{aligned}}
\newcommand{\beq}{\begin{equation}}
\newcommand{\eeq}{\end{equation}}
\newcommand{\nn}{\nonumber}

\def\CP{{{\mathbb C}{\mathbb P}}}
\def\0{{\sst{(0)}}}
\def\1{{\sst{(1)}}}
\def\2{{\sst{(2)}}}
\def\3{{\sst{(3)}}}
\def\4{{\sst{(4)}}}
\def\5{{\sst{(5)}}}
\def\6{{\sst{(6)}}}
\def\7{{\sst{(7)}}}
\def\8{{\sst{(8)}}}
\def\sst#1{{\scriptscriptstyle #1}}
\def\oneone{\rlap 1\mkern4mu{\rm l}}
\def\ep{{\epsilon}}
\def\del{{\partial}}

\def\dfft#1#2{{\displaystyle\fft{#1}{#2}}}

\def\im{{{\rm i\,}}}
\def\R{{\mathbb R}}

\def\Z{{\mathbb Z}}
\def\C{{\mathbb C}}
\def\Q{{\mathbb Q}}
\def\alp{{{\alpha}}}
\def\bet{{{\beta}}}

\newcommand{\overbar}[1]{\mkern 1.5mu\overline{\mkern-1.5mu#1\mkern-1.5mu}\mkern 1.5mu}

\def\wtd{\widetilde}

\def\ie{{i.e.~}}

\def\ii{{\rm i}}

\def\ben{\begin{equation}}
\def\bea{\begin{eqnarray}}
\def\een{\end{equation}}
\def\eea{\end{eqnarray}}

\def\ft#1#2{{\textstyle{\frac{\scriptstyle #1}{\scriptstyle #2} } }}
\def\fft#1#2{{\frac{#1}{#2}}}
\newcommand{\eq}[1]{(\ref{#1})}
\newcommand{\hatslashed}[1]{{\hat{#1}\llap{/\kern0.2em}}}

\title{\boldmath Wrapping M5-branes on a non-spin manifold: Spin$^h$ structures and AdS$_7$ black holes}

\author[a]{Cameron Gibson,}

\author[a]{Okan G\"unel,}

\author[a,b]{and C.N. Pope}

\affiliation[a]{George P. \& Cynthia W. Mitchell Institute for Fundamental Physics \& Astronomy,\\
         Texas A\&M University,
         College Station,
         TX 77843-4242,
         USA.}
\affiliation[b]{DAMTP, Centre for Mathematical Sciences,\\
Cambridge University,
Wilberforce Road, Cambridge CB3 OWA, UK}

\emailAdd{camerongibson@tamu.edu, okan.gunel@tamu.edu, pope@physics.tamu.edu}

\abstract{We construct a solution of seven-dimensional maximal gauged supergravity of the form AdS$_2\times M_5$, where $M_5$ is the five-dimensional Wu manifold $SU(3)/SO(3)$.  The solution is supersymmetric, despite the fact that the Wu manifold admits neither a spin structure nor a spin$^c$ structure. Rather, it admits a spin$^h$ structure, wherein the ill-definedness of ordinary spinors due to a minus sign in their holonomy is compensated by a further minus sign arising through the spinors carrying $SO(3)$ Yang-Mills charge, in a half-integer isospin representation of the $SU(2)$ double cover of the $SO(3)$.  We show how the configuration can be lifted to a solution of $D=11$ supergravity of the form AdS$_2\times M_9$, where
$M_9$ is the unit tangent bundle of the Wu manifold.  We discuss the geometry and topology of $M_9$, and study the supersymmetry from the eleven-dimensional standpoint. We extend the seven-dimensional solution to a black hole, and discuss some aspects of its holographic interpretation and the UV to IR flow.  We revisit a previously-known AdS$_2\times H^3\times H^2$ solution, and its UV to IR flow, showing that the flow equations in fact admit of an exact solution. Additionally, we give a detailed discussion of the lift of the AdS$_2\times S^5$ solution to $D=11$, exhibiting similarities and differences
between the properties of its internal space $M_9$ (the unit tangent bundle of $S^5$) and the Wu manifold example of our main discussion.  We also construct the non-compact dual of the Wu manifold solution, where $M_5$ is taken to be $SL(3,\R)/SO(3)$.
}

\begin{document}
\maketitle
\flushbottom

\section{Introduction}

The $SU(3)/SO(3)$ coset manifold, known as the Wu manifold, has some interesting 
properties, one of which is that it is neither a spin nor a spin$^c$ manifold \cite{notspinc}, 
but rather a spin$^h$ manifold \cite{Chen:2017}. This means that for spinors to be globally defined, 
they must carry charge under  
at least an $SO(3)$ non-abelian gauge group. In a recent paper \cite{Wu}, 
the local properties and the obstruction to the spin structure were shown geometrically for the Wu manifold.
This manifold supports a natural $SO(3)$ Yang-Mills field.   
The obstruction to the existence of a spin structure on the Wu manifold arises because if uncharged spinors are introduced on the manifold, the holonomy 
around a non-trivial 2-cycle gives a minus-sign inconsistency.  This can be overcome by taking the spinors 
to be charged under the $SO(3)$ gauge group, but in a half-integer representation of the $SU(2)$ covering group.
This introduces a further minus sign in the holonomy, which compensates the original one arising for
uncharged spinors \cite{hawpop}, thus allowing the charged spinors to be defined globally on the Wu manifold.

Five is the lowest dimension in which an orientable manifold that is not spin or spin$^c$ can arise, and so the
Wu manifold can be viewed as providing the canonical example:  All
orientable manifolds of dimension $D\le 3$ are spin, while in $D=4$ all orientable manifolds are either 
spin or spin$^c$ \cite{HirzebruchHopf1958} ($\CP^2$ being the canonical example of the latter).  Only when $D\ge 5$ can the
third Stiefel-Whitney class $w_3$ be non-vanishing, signalling the absence of a spin$^c$ structure. 

There are only two compact simply-connected irreducible symmetric spaces in five dimensions, namely $S^5$ and the Wu manifold.  The Wu manifold is in fact a rational homology 5-sphere; its rational homology groups $H_i(W,\Q)$ are the same as those of the 5-sphere, $H_i(S^5;\Q)=(\Q,0,0,0,0,\Q)$. Furthermore, five is the lowest dimension where a rational homology sphere is not actually a sphere (as we shall discuss later, the Wu manifold has torsion in its integral homology, which $S^5$ does not). 
The Wu manifold also plays an important role as a generator of the oriented cobordism group $\Omega_5^{SO}\cong \Z_2$ \cite{crowley11}. It was used in \cite{debray2} to show the triviality of a potential nonperturbative anomaly of the $E_{7(7)}(\R)$ $U$-duality symmetry of $4d$ ${\cal N}=8$ supergravity.
Thus from the standpoint of supergravity compactifications, the Wu manifold enjoys a rather significant status as a companion and a contrast to the 5-sphere.  It would not provide an interesting alternative to the 5-sphere in an application such as the standard  compactification of type IIB supergravity, because of its lack of a spin structure.  However, as we shall see, in other contexts its spin$^h$ structure can be exploited so as to give a satisfactory supergravity compactification with fermions and indeed supersymmetry.

Non-spin manifolds are often avoided in the supergravity compactifications,
owing to the problems of not having globally defined spinors or Killing spinors. However, in this 
paper, we obtain a supersymmetric solution of $D=7$ gauged maximal supergravity \cite{Pernici:1984xx, Nastase:1999cb, Nastase:1999kf} that uses the Wu manifold 
in the internal space. By embedding the non-abelian $SO(3)$ gauge fields of 
the Wu manifold in the $SO(5)$ gauge group of the seven-dimensional theory, we obtain
a seven-dimensional solution of the form AdS$_2 \times W$, where $W$ is the Wu manifold, with globally-defined spin$^h$
spinors. This can be uplifted automatically to a $D=11$ supergravity solution since the $D=7$ theory arises as a consistent $S^4$ reduction from the $D=11$ theory. Furthermore, as discussed in detail in \cite{Wu}, the Wu manifold admits a gauge-covariantly
constant spinor (in the spin-$\ft32$ representation of the $SU(2)$ double cover of $SO(3)$), and we show that
this implies that the AdS$_2 \times W$ solution of seven-dimensional gauged supergravity, and hence also of $D=11$ supergravity, is supersymmetric. 

We show also that the AdS$_2\times W$ solution can be generalised to give a non-extremal black hole solution of the gauged seven-dimensional supergravity, which becomes supersymmetric in the extremal limit.  The previously-mentioned 
AdS$_2\times W$ solution can be understood as the near-horizon limit of the extremal black hole. 

In \cite{Gauntlett:2000ng, Gauntlett:2001jj, Gauntlett:2002rv}, similar supersymmetric AdS$_2\times M_5$ solutions of 7D gauged maximal supergravity were found, where $M_5$ is $S^5$, $H^5$, or $H^3 \times H^2$. Here, we shall use a similar ansatz but with $M_5$ being the $SU(3)/SO(3)$ Wu manifold and its noncompact dual $SL(3,\R)/SO(3)$. 

The paper is structured as follows. 
In section~\ref{sec:2}, we recall the basics of the gauged maximal 7D supergravity \cite{Pernici:1984xx}, which we will use as our parent theory. We review the field content, bosonic action, equations of motion, and the supersymmetry transformations of the fermionic fields in the theory. 
Then, in section \ref{sec:3}, we present an ansatz for solving the gauged maximal 7D supergravity. The ansatz starts with a metric in the form of AdS$_2\times$Wu manifold and it is similar to the ansatz of \cite{Gauntlett:2000ng}. Then, we embed $SO(3)$ in $SO(5)_g$ gauge group of the 7D theory via the principal embedding. This allows us to use the $SO(3)$-gauge covariantly constant spinor of the Wu manifold and its special properties \cite{Wu} in the gravitino and dilatino ans\"atze. 
The scalar fields of the $D=7$ supergravity are zero in the compactification, which allows us to identify the $SO(5)_g$ gauge group indices with the local composite $SO(5)_c$ under which the fermions transform. The 2-form field strengths are chosen to be the $SO(3)$ gauge fields on the Wu manifold, and we use these to solve for the 3-form potentials. Then, we check the bosonic equations of motion. The Einstein equations give constraints on the ansatz, and later, we obtain the same constraints from the variation of the fermionic fields. Thus, we conclude that our solution is supersymmetric. We also extend the setup to the noncompact dual of the Wu manifold, the $SL(3,\R)/SO(3)$ coset, and find also a supersymmetric AdS$_2\times SL(3,\R)/SO(3)$ solution. 

In section \ref{D11sec} we lift the solution to eleven-dimensional supergravity, using the results on the consistent reduction on $S^4$ that were obtained in \cite{Nastase:1999cb,Nastase:1999kf}.  The solution in $D=11$ is then of the form AdS$_2\times M_9$, where $M_9$ is an $S^4$ bundle over the five-dimensional Wu manifold. This overcomes the inability to construct a solution using a direct product of the Wu manifold and $S^4$ \cite{fof}. We discuss some properties of $M_9$ in detail, showing that the metric takes the form of an inhomogeneous metric of cohomogeneity 1, whose principal orbits are 
the homogeneous space $SU(3)/(\Z_2\times \Z_2)$ and it admits an $SU(4)$ structure.  This makes contact with some general results on supersymmetric $D=11$ supergravity compactifications of the form AdS$_2\times M_9$ in \cite{zayas}.  Our specific construction based on the Wu 
manifold provides a new and especially interesting example of this class of compactifications within the more general work to classify AdS$_2$ solutions of supergravity \cite{class1,class2,class3,class4}.

In section \ref{sec:4}, we generalise the AdS$_2\times W$ solution of $D=7$ gauged maximal supergravity to an extremal black hole. By adding a mass parameter to the solution, we obtain a Schwarzschild-AdS$_7$ type of black hole with $SO(3)$ Yang-Mills fields and a 3-form gauge potential where the $S^5$ in AdS$_7$ is replaced with the Wu manifold. We observe that the first solution in section \ref{sec:4} is its extremal limit and the solution in section \ref{sec:3} becomes its near-horizon limit.
We continue by calculating thermodynamic quantities of the solution and conclude the section with remarks on the holographic applications. 
Finally, we present some concluding remarks in section \ref{sec:5}.

Five appendices contain further details of some formalism and calculations.  In appendix \ref{wurecap}, we record some useful expressions for the metric on the Wu manifold $SU(3)/SO(3)$, and its $SO(3)$ Yang-Mills gauge fields.  Appendix \ref{kkcalc} gives some standard formulae for the Kaluza-Klein lift on $S^4$ of the seven-dimensional metric to eleven dimensions. The solutions we are considering then take the form AdS$_2\times M_9$, where $M_9$ is the unit tangent bundle of the original five-dimensional internal space $M_5$ of the seven-dimensional solution.  In appendix \ref{pz}, the equations governing the $SU(4)$ structure of the manifold $M_9$, which we make use of in the paper, are reproduced from \cite{zayas}. Appendix \ref{S4overS^5} contains a detailed discussion of the lift of the AdS$_2\times S^5$ solution of \cite{Gauntlett:2000ng}. We discuss some geometric and topological aspects of the manifold $M_9$ (which in this case is the homogeneous Stiefel manifold $SO(6)/SO(4)$), and we give a detailed discussion of the supersymmetry of this background.  Appendix \ref{apph2h3} discussed one of the other seven-dimensional AdS$_2\times M_5$ solutions considered in \cite{Gauntlett:2001jj},
where $M_5=H^3\times H^2$. We extend the results of \cite{Gauntlett:2001jj} by constructing an exact solution for the flow equations, which had previously only been studied numerically.

\bigskip
\noindent{\bf Notation and Conventions:}
\medskip

We conclude the introduction with some notation and conventions that we shall be using in the paper.

The indices $M, N$ will be used to denote tangent-frame indices on AdS$_2\times W$, and these will be 
decomposed as $M=(\alpha,a)$, 
with $\alpha, \beta=0,1$ denoting AdS$_2$ tangent-frame indices and $a, b=1,\dots,5$ denoting Wu manifold tangent-frame indices. The indices $A, B=1,\dots,5$ denote $SO(5)_g$ gauge indices, $i, j=1,\dots,5$ denote $SO(5)_c$ composite indices and $I, J=1, \dots,4$ denote $SO(5)_c$ spinor indices.  Note that the vector indices of $SO(5)_g$ and $SO(5)_c$ can be freely raised and lowered using a Kronecker delta. We shall use the indices $i, j=1,2,3$ again for the $SO(3)$ indices when they are placed on the $SO(3)$ gauge potential, field strength, and generators on the Wu manifold.

The 7D spacetime gamma matrices are decomposed as
\begin{equation}
  \hat{\gamma}_M =
    \begin{cases}
      \td\gamma_\alpha \otimes I_4 & M=\alpha\in \text{AdS}_2\\
      \td\gamma_\ast\otimes\gamma_a & M=a\in\text{Wu}\\
    \end{cases}       \, ,
    \label{g7d}
\end{equation}
where the 2D gamma matrices $\td\gamma_\alpha$ are such that the chirality operator is
\begin{equation}
    \td\gamma_\ast=\frac{1}{2}\varepsilon_{\alpha\beta}\,\td\gamma^{\alpha\beta}\, ,
    \label{gamma3}
\end{equation}
and $\gamma_a$ are the 5D gamma matrices
\begin{equation}
\gamma_1=\tau_2 \otimes \tau_1 \, , \quad 
\gamma_2=\tau_2\otimes \tau_2 \, , \quad
\gamma_3=\tau_2\otimes \tau_3 \, , \quad
\gamma_4=-\tau_1\otimes I_2 \, , \quad
\gamma_5=\tau_3 \otimes I_2 \, ,\label{gammadef}
\end{equation}
where $\tau_i$ are the Pauli matrices.

The gamma matrices associated with $SO(5)_c$ are
\begin{equation}
\begin{split}
\Gamma_1=\tau_3 \otimes \tau_2\, , \quad 
\Gamma_2=\tau_3\otimes \tau_1 \, , \quad
\Gamma_3=-\tau_2\otimes I_2 \, , \quad
\Gamma_4=-\tau_1\otimes I_2\, , \quad
\Gamma_5=\tau_3 \otimes \tau_3 \, .
\label{gaugegamma}
\end{split}
\end{equation}

We shall choose $\varepsilon_{\alpha \beta}$ to be such that
\begin{equation}
    \varepsilon_{01}=+1\,,
\end{equation}
and $\varepsilon^{\alp} {}_\beta=\bar{\eta}^{\alp \delta} \varepsilon_{\delta \beta}$ where $\bar{\eta}=\text{diag$(-1,1)$}$.

\section{The \texorpdfstring{$SO(5)$}{SO(5)} gauged maximal 7D supergravity}
\label{sec:2}

\subsection{The field content}
The gauged maximal $D=7$ supergravity \cite{Pernici:1984xx} is obtained by the consistent reduction of $D=11$ supergravity on $S^4$ \cite{Nastase:1999kf, Nastase:1999cb}, and it has a local gauge group $SO(5)_g$. 
The scalars parametrise the coset $SL(5,\R)/SO(5)_c$ and transform under both $SO(5)_g$ and $SO(5)_c$. The subscript $c$ in $SO(5)_c$ stands for composite. The field content under $SO(5)_g$ and $SO(5)_c$ is given in table \ref{tab:1}.
\begin{table}[h!]
\centering
\begin{tabular}{lccccc}
\toprule
 & $\psi_{M}^{I}$ &  $B_{M}^{AB}$ & $S_{MNR,A}$ & $\lambda^{I}_{i}$ & $\Pi_A^i$ \\
\midrule
$SO(5)_g$ & $\mathbf{1}$ & $\mathbf{10}$ & $\mathbf{5}$ & $\mathbf{1}$  & $\mathbf{5}$ \\
$SO(5)_c$ & $\mathbf{4}$ & $\mathbf{1}$ & $\mathbf{1}$ & $\mathbf{16}$ & $\mathbf{5}$ \\
\bottomrule
\end{tabular}
\caption{\label{tab:1}The field content of the 7-dimensional gauged maximal theory under $SO(5)_c$ and $SO(5)_g$.}
\end{table}

The field content includes the vielbein $e$, the 4 gravitini $\psi_M^{I}$, the $SO(5)_g$ vector potentials $B_M ^{AB}$, the 3-form potentials $S_{MNP , A}= S_{[MNP], A}$, the spin $\ft12$ fields $\lambda_i {}^{I}$ and the scalars $\Pi^i_A$. We shall suppress the $SO(5)_c$ spinor indices $I,J,\cdots$ in the rest of the paper.

\subsection{Bosonic fields}
The bosonic part of the Lagrangian \cite{Pernici:1984xx} is
\bea
&2{\cal L}=e\left[R+\frac{1}{2}m^2(T^2-2T_{ij}T^{ij})-
P_{M ij}P^{M ij}-\frac{1}{2}({\Pi_A}^i{\Pi_B}^j
G_{MN}^{AB})^2-m^2({{\Pi^{-1}}_i}^A S_{MNP,A})^2\right]\nn \\
&-{6m}\,S_A\wedge F_{(4)}^A +{\sqrt 3}
\epsilon_{ABCDE}\delta^{AF}S_F\wedge G^{BC}\wedge
G^{DE}
+\fft1{8m}\,(2\Omega_5[B]-\Omega_3[B]) \,  ,
\eea
where $G^{AB}$ is the $SO(5)_g$ Yang-Mills field strength and $F_{(4)}^A=F_{(4)B} \delta_A^B$ is the four-form field strength defined by
\begin{equation}
    F_{(4)}^A=dS^A+g\, B^A{}_B\wedge S^B \, .
\end{equation}
$\Omega_5[B]$ and $\Omega_3[B]$ are the Chern-Simons term for the gauge potential $B$.
$P_{M,ij}$ is associated with the scalar kinetic term and defined by
\beq
\Pi^{-1} {}_i {}^A (\delta_{A}^B \, \partial_M + g \, B_{M,A} {}^B ) \Pi_B {}^k \delta_{kj} 
= Q_{M,[ij]} + P_{M,(ij)} \,\, ,
\eeq
where $Q_{M,ij}$ is the $SO(5)_c$ composite gauge field.
The scalar potential is
\begin{equation}
e^{-1}V =- \frac{1}{4} m^2 (T^2 - 2T_{AB}T^{AB})\, ,
\end{equation}
where $T_{AB} = \Pi^{-1\ i}_A \,\Pi^{-1\ j}_B \delta_{ij}$ and $T = T_{AB}\delta^{AB}$.

The gauge covariant derivative acts on an object which transforms under both $SO(5)_g$ and $SO(5)_c$ (indices $A,B,\ldots$ and $i,j,\ldots$, respectively) as
\beq
D_M V_{Ai} = \nabla_M V_{Ai} + g B_{M,A} {}^B V_{Bi} + Q_{M, i} {}^j V_{Aj} \, , \label{eq:gaugecovder}
\eeq
where $g=2m$.

The bosonic equations of motion that will be relevant in our discussion below comprise the Einstein field equations
\bea\label{einst}
R_{MN}&=&P_{M}P_{N}+
(\Pi \,\Pi G)_{MP}(\Pi \,\Pi G)_N {}^P
+3m^2({\Pi^{-1}}S)_{MPQ}
(\Pi^{-1} S)_{N}{}^{PQ}\nn \\
&&-\frac{1}{10} g_{MN}\left[m^2(T^2-2T_{ij}T^{ij})
+({\Pi}\,{\Pi} G)^2
+4m^2(\Pi^{-1}S)^2\right]\,,
\eea  
where $({\Pi}\,{\Pi} G)^2$ is shorthand for $(\Pi^i_A\, \Pi^j_B\, G^{AB}_{MN})^2$, etc., 
and the equations of motion for the 3-form potentials $S^A$, which are
\beq
6 m^2 \ast_7 S^A = -12 m F_{(4)}^A - \sqrt{3}\, \ep_{ABCDE} \, G^{BC} \wedge G^{DE}\,.
\label{Seom}
\eeq
The Yang-Mills equations for the fields $G^{AB}$ will be automatically satisfied as a consequence of the fact that they will be expressed in terms of $SO(3)$ gauge fields on the five-dimensional Wu manifold that obey the five-dimensional Yang-Mills equations.  The scalar fields $\Pi^i_A$ will be trivial in the backgrounds we shall consider, i.e. $\Pi^i_A=\delta^i_A$, 
and their equations of motion will be automatically satisfied.

\subsection{Fermionic fields}

We shall be concerned with establishing the supersymmetry of the background that we construct, and so we shall need the supersymmetry transformations of the fermionic fields.  These are given, up to the order relevant for 
establishing the supersymmetry of a bosonic background, by the
gravitino transformation rule (isospin and spin indices are suppressed)
\beq
\bald
\delta\psi_{M} = D_{M}\epsilon + \frac{1}{4}m\,\hat{\gamma}_{M}\epsilon 
& - \frac{1}{40}({\hat{\gamma}_{M}} {}^{NP} 
- 8\delta_{M} {}^{N}\hat{\gamma}^{P})\Gamma_{ij} \Pi_{A} {}^i \Pi_{B} {}^j G_{NP}^{AB} \, \epsilon  \\
&+\frac{m}{10 \sqrt{3}} \left(\hat{\gamma}_M {}^{NPR}- \frac{9}{2} \delta_M {}^N \hat{\gamma}^{PR} \right) \Gamma^i \Pi^{-1}_i {}^A S_{NPR,A}  \, \epsilon \,,
\label{eq:gravvar}
\eald
\end{equation}
and the dilatino transformation rule
\begin{align}
\delta\lambda_i=
   \frac{1}{2}\hat{\gamma}^{M}\Gamma^j  P_{M ij} \,\epsilon
   &+ \frac{1}{2}m(T_{ij}-\frac{1}{5}T\delta_{ij})\Gamma^j\epsilon
   + \frac{1}{16}\hat{\gamma}^{MN}(\Gamma_{kl}\Gamma_i
       -\frac{1}{5}\Gamma_i\Gamma_{kl})
        {\Pi_A}^k{\Pi_B}^l G_{MN}^{AB}  \, \epsilon \nonumber\\
   & + \frac{m}{20\sqrt{3}}\hat{\gamma}^{MNP}(\Gamma_i{}^j-4\delta_i{}^j)
       {{\Pi^{-1}}_j}^A S_{MNP ,A} \, \epsilon
       \label{fulldilvar} \, .
\end{align}

\section{7D gauged supergravity solution}
\label{sec:3}

We decompose the 7D spacetime into AdS$_2 \times W$ where $W$ is the Wu manifold by making the ansatz
\begin{equation}
ds_7^2 = ds_{\text{AdS}_2}^2(x) + \lambda^2 ds_\text{Wu}^2(y) \, ,
\end{equation}
where $ds_\text{Wu}^2$ is the canonically-normalised Wu metric ($\bar{R}_{ab} = 6\delta_{ab}$) as given in \cite{Wu} and briefly recapped in appendix~\ref{wurecap}, $ \lambda$ parametrises the internal scale size, and $ds^2_{\text{AdS}_2}$ is the metric on 2-dimensional AdS external space with $\bar{R}_{\alp \bet}=\Lambda \bar\eta_{\alp \bet}$. We also define 
\bea
C=\fft1{m\lambda}\label{Cdef}
\eea
for later convenience. 

Our aim is to use the gauge-covariantly constant spinor on the Wu manifold, which is in the $j=\ft32$ representation of the $SU(2)$ double cover of $SO(3)$ \cite{Wu}, to find supersymmetric solutions. Thus, we are going to embed the structural $SO(3) \hookrightarrow SO(5)_g$ via the principal embedding. Under this embedding, the spinor $\mathbf{4}$ of $SO(5)$ does not branch, but transforms as the irreducible $\mathbf{4}$ of the double cover of $SO(3)$, that is, isospin $j=\ft32$\footnote{
Although there are three distinct embeddings of $SO(3)$ into $SO(5)$, the principal embedding we consider is the only one that can lead to a supersymmetric solution. This is because the only gauge-covariantly constant spinor satisfying $D_M \ep=0$ (which needs to hold to make the gravitino variation in eqn \ref{eq:gravvar} vanish) transforms in the $\mathbf{4}$ irreducible spinor representation of $SO(3)$. The other two embeddings do not lead to spinors charged as $j=\frac{3}{2}$ and so cannot lead to a supersymmetric solution, although this does not rule out the possibility of other bosonic solutions.}
In order to identify the 7D $SO(5)_g$ Yang-Mills connection $B_M^{AB}$ with the $SO(3)$ connection $A_a^i$ of the Wu manifold, we use
\begin{equation}
B_a^{AB} = \frac{1}{2m \lambda} A_a^{AB} \equiv\frac{1}{2m \lambda} A_a^i (T^i)^{AB} \, ,
\label{WuPotEmbed}
\end{equation}
where $1/(2m)$ factor has been chosen according to the eqn \eq{eq:gaugecovder} and $g=2m$, such that the gauge covariant derivative is the one in the Wu manifold. Note that $A^i_a$ is the gauge potential on the canonically-normalised Wu metric $ds_\text{Wu}^2$ and explicit expressions for these gauge potentials and corresponding field strengths are given in appendix~\ref{wurecap}. The 7D field strengths $G_{ab}^{AB}$ are then related to the field strengths on the Wu manifold $F_{ab}^i$ by
\begin{equation}
G_{ab}^{AB} = \frac{1}{2m \lambda^2} F_{ab}^{AB} \equiv \frac{1}{2m \lambda^2} F_{ab}^i (T^i)^{AB} \, .
\label{WuFieldEmbed}
\end{equation}
The $SO(3)$ subset $T^i$ of the $SO(5)_g$ generators in the $5$-dimensional representation, with components $T^i_{AB}$, can be given by the components of the $SO(3)$ gauge field strengths $F^i_{ab}$ in the Wu
manifold:
\beq
(T^i)_{AB} = F^i {}_{AB} \, .
\label{Teq}
\eeq
Note that this means we are identifying the $SO(5)_g$ group with the $SO(5)$ tangent-frame group of the Wu manifold, and so the $A,B,\ldots$ $SO(5)_g$ indices become interchangeable with the $a,b,\ldots$ $SO(5)$ tangent-frame indices.
(Note that the field strengths $F^i_{ab}$ on the Wu manifold obey the $SO(3)$ algebra $[F^i,F^j]_{ab}\equiv 
F^i_{ac}\, F^j_{cb}-
 F^j_{ac}\, F^i_{cb} = \epsilon_{ijk}\, F^k_{ab}$ \cite{Wu}.)

We fix the scalars at the trivial coset origin:  $\Pi^i_A = \delta^i_A$. This gives 
\beq
P_M^{ij}=0 \, , \quad Q^{ij}_M=g B^{ij}_M \, , \quad T_{AB} = \delta_{AB}  \quad \text{and} \quad  T=5 \, . 
\eeq
Hence, the scalar potential becomes
\beq
e^{-1}V=-\frac{15}{4}m^2 \, .
\eeq
Note that imposing $\Pi^i_A = \delta^i_A$ implies that the $SO(5)_c$ indices $i,j,\ldots$ and the $SO(5)_g$ indices $A,B,\ldots$ become interchangeable, in addition to the equivalence of the $SO(5)_g$ indices and the five-dimensional
indices of the Wu tangent frame that we saw above.

The spinor gauge covariant derivative is defined by
\begin{equation}
    D \equiv \nabla+\frac{1}{4}Q_{ij}\Gamma^{ij}\,,
\end{equation}
where $\nabla= d+\frac{1}{4}\omega _{ab}\gamma^{ab}$
is the usual spinor covariant exterior derivative.

By fixing the scalars at the origin, composite $SO(5)_c$ indices become identified with $SO(5)_g$ indices. Taking $Q^{ij}=gB^{ij}$ then gives
\begin{equation}
\begin{split}
    D=&\nabla+\frac{g}{4}B_{AB}\Gamma^{AB}\\
    =&\nabla+\frac{1}{4}A_{AB}\Gamma^{AB}\\
    =&\nabla+\frac{1}{4}A^i(T^i)_{AB}\Gamma^{AB}\\
    =&\nabla+\frac{1}{4}A^i(F^i)_{AB}\Gamma^{AB}\\
    =&\nabla-\ii \, \hat{T}^iA^i \, ,
\end{split}
\end{equation}
where $\hat{T}^i \equiv \frac{\ii}{4}\Gamma_{AB}(T^i)^{AB}$ are the $j=\frac{3}{2}$ generators of the $SU(2)$ double cover of $SO(3)$\footnote{When we refer to an $SO(3)$-gauge covariantly constant spinor, we actually mean a charged spin$^h$ spinor in the sense that fermions are charged with half-integer isospin $j$, analogous to how spin$^c$ spinors on $\CP^2$ are charged with half the usual $U(1)$ charge that Dirac quantisation requires. However, bosons are charged with the usual integer values of $j$. In other words, the $SO(3)$ gauge bundle on the Wu manifold only appears with half-integer $j$ when tensored with the spinor bundle. An in-depth discussion of this on the Wu manifold and $\CP^2$ is given in \cite{Wu}\label{footnote}.} With the choice of gamma matrices \eq{gaugegamma}, and eqn \eq{Teq}, $\hat{T}^i$ become the generators $T^i$ used to generate the gauge-covariantly constant spinor $\eta$ in section 4.6 of \cite{Wu}. Thus, the $SO(3)$-gauge covariantly constant spinor $\eta$ on the Wu manifold satisfies $D \eta = 0$ \cite{Wu}.

We decompose the 7D spinor $\epsilon$ as 
\beq
\epsilon(x,y) = \xi(x) \otimes \eta(y) \, ,
\label{spinansatz}
\eeq
where $\eta(y)$ is the gauge-covariantly constant spinor on the Wu manifold and $\xi(x)$ is the Killing spinor on AdS$_2$ such that
\begin{equation}
    \nabla_\alpha \xi = c \, \td\gamma_\alpha\, \xi \, ,
\end{equation}
for $\Lambda=-4c^2$.
Note that the suppressed $SO(5)_g$ spinor index on $\epsilon$ in eqn (\ref{spinansatz}) 
is carried by $\eta$ on the right-hand side. 

Finally, the gauge-covariantly constant spinor $\eta$ (also $\epsilon$) on the Wu manifold obeys the following identities that we use frequently:
\bea
    F^{AB}_{ab}\gamma^{ab}\Gamma_{AB}\, \eta &=&{F^{iAB}}F^i_{MN}\gamma^{ab}\Gamma_{AB}\eta=60\, \eta \, ,
        \label{Fident} \\
    F^{AB}_{ab}\gamma^{b}\Gamma_{AB}\, \eta &=&-12 \gamma_a \, \eta \, ,
    \label{Fident1} \\
    F^{AB}_{ab}\gamma^{ab}\Gamma_{B}\, \eta & =& 12 \Gamma^A\,\eta \, ,
    \label{Fident2} \\
    (I_4 \otimes\Gamma_A )\, \eta &=& \delta^a_A(\gamma_a\otimes I_4)\,  \eta \, ,
    \label{gdiden} \\
    (I_4 \otimes\Gamma_{AB} )\, \eta &=&
    -\delta^a_A\delta^b_B(\gamma_{ab}\otimes I_4)\,  \eta \, .
    \label{gddiden}
\eea

\subsection{3-form potential}
In order to satisfy eqn \eq{Seom}, we take the same ansatz that is taken for five-cycles in 
\cite{Gauntlett:2000ng}, namely
\beq
S^A=a \, \delta_a^A\, \epsilon_2 \wedge e^a \, ,
\label{Sanz}
\eeq
where $\epsilon_2$ is the volume form of AdS$_2$. It has a vanishing 4-form field strength $F_{(4)}$:
\bea
F_{(4)}^A & = & D S^A \, \equiv \,  d S^A + g B^{A}{}_B \wedge S^B \nn \\
& = & a \, \epsilon_2 \wedge(\delta_a^A \, d e^a +  \delta_b^B \, (F^i)^A {}_B A^i \wedge e^b ) \nn \\
& = & a \,\delta_a^A  \,\epsilon_2 \wedge( d e^a +  w^a {}_b \wedge e^b )=0 \nn 
\eea
where in the second line, we used eqns \eq{WuPotEmbed} and \eq{Teq}, in the third line we used $w^a {}_b=F^{ia} {}_bA^i$ from \cite{Wu}, and the result follows from the Maurer-Cartan equations on the Wu manifold. 

Plugging eqn \eq{Sanz} in the equation of motion \eq{Seom}, we find that the $S_{MNP,A}=S_{[MNP],A}$ are specified by the non-vanishing components 
\begin{equation}
    S_{\alpha \beta a,A}=-2\sqrt{3}\,C^4\, \varepsilon_{\alpha \beta}\,\delta_{aA} \,,
    \label{Ssol}
\end{equation}
where $C$ is defined in eqn (\ref{Cdef}).

\subsection{Einstein equations}
\label{sec:Eins}

In this section, we use our ansatz in the Einstein equation \eq{einst} to determine the internal radius and the cosmological constant of the AdS$_2$ space. 
The Einstein equations simplify to
\bea\label{einst2}
R_{MN}=
 G_{MP} G_N {}^P
+3m^2S_{MPQ}
 S_{N}{}^{PQ} -\frac{1}{10} g_{MN}\left(15m^2
+G^2
+4m^2S^2\right) \, .
\eea  
Let us start with the terms in parentheses since they will appear in both the internal and external directions:
\beq
\bald
G^2 =& G_{ab}^{AB} G_{ab}^{AB} = \frac{1}{4m^2 \lambda^4} F_{ab}^{AB} F^{ab}_{AB} \\
=&  \frac{1}{4m^2 \lambda^4} F^i_{ab} F^{iAB} F^{jab} F^j_{AB} =  \frac{100}{4m^2 \lambda^4} \delta^i_i =  \frac{75}{m^2 \lambda^4} \, , \\
S^2= & S_{MNP,A} S^{MNP,A} = \frac{36}{m^8 \lambda^8} \varepsilon_{\alp \beta} \varepsilon^{\alp \beta} \delta_{aA} \delta^{aA} \\
=& -\frac{360}{m^8 \lambda^8} \, .
\eald
\eeq

\subsubsection{Internal directions}
In the internal directions, we can use $R_{ab}=\frac{1}{\lambda^2}\bar{R}_{ab}= \frac{6}{\lambda^2}\delta_{ab}$. As stated earlier, we will use $C=(m\lambda)^{-1}$ for convenience. Thus, the Einstein equations become
\beq
\bald
6\delta_{ab} &= \frac{C^2}{4} F^{AB}_{ac}F_{bc}^{AB}
+36C^6 \varepsilon_{\alp \bet} \varepsilon^{\alp \bet} \delta_{ab} 
 - \frac{1}{10} \delta_{ab} \left(\frac{15}{C^2} + 75C^2-1440C^6 \right) \\
 &= \frac{60C^2}{4} \delta_{ab}  
-72 C^6  \delta_{ab} 
 - \frac{1}{10} \delta_{ab} \left(\frac{15}{C^2} + 75C^2-1440 C^6 \right) \\
  &= \left(- \frac{3}{2C^2} + \frac{15}{2}C^2 + 72C^6 \right) \delta_{ab}   \, .
 \eald
 \eeq
This simplifies to the algebraic relation
\beq
1 + 4 C^2 -5 C^4 - 48C^8 =0 \, ,
\eeq
which factorises into 
\beq
\left( 1 + 3 C^2 -12C^4 \right) \left( 1 + C^2 + 4C^4 \right) = 0 \, .
\label{einsint}
\eeq
The first factor, which admits a real solution for $C$ with $C^2=\fft{\sqrt{57}+3}{24}$,  
will also be encountered in the gravitino supersymmetry variation, which we shall study below. Rather remarkably, despite the rather unpromising-looking occurrence of the quadratic surd in the expression for the parameter $C^2$, it will turn out that the AdS$_2\times W$ solution is in fact supersymmetric.

\subsubsection{External directions}

In the AdS$_2$ directions, $R_{\alp \bet}=\Lambda \bar{\eta}_{\alp \bet}$, and $G_{\alp M} G_{\bet} {}^{M} = 0$ since the $SO(3)$ gauge potential lives only in the Wu directions. Hence, the Einstein equations in the AdS$_2$ directions become
\bea
\Lambda \lambda^2 \,  \bar{\eta}_{\alp \bet} &=& 72C^6 \varepsilon_{\alp \gamma} \,\varepsilon_{\beta \delta} \,\bar{\eta}^{\gamma \delta} \, \delta_{a}^a 
 - \frac{1}{10} \bar{\eta}_{\alp \bet} \left(\frac{15}{C^2} + 75C^2-1440C^6 \right) \\
  &=& \left(- \frac{3}{2C^2} - \frac{15}{2}C^2 - 216C^6 \right) \bar{\eta}_{\alp \bet}  \, .
  \eea
We will see later that the $\Lambda$ found from considerations of supersymmetry in section \ref{sec:gravext} is consistent with this equation.

\subsection{Dilatino variation}
With our ansatz, the dilatino variation eqn \eq{fulldilvar} simplifies to
\begin{align}
\delta\lambda_i=
    \frac{1}{16}\hat{\gamma}^{MN}(\Gamma_{kl}\Gamma_i
       -\frac{1}{5}\Gamma_i\Gamma_{kl})
       \, G_{MN}^{kl} \, \epsilon  + \frac{m}{20\sqrt{3}}\hat{\gamma}^{MNP}(\Gamma_i{}^j-4\delta_i{}^j)
      \, S_{MNP ,j} \, \epsilon \, .
\end{align}
Substituting $S$ and $G$ into eqn \eq{g7d} gives
\beq
\delta\lambda_i=
    \frac{1}{32m\lambda^2}  \gamma^{ab}(\Gamma_{kl}\Gamma_i
       -\frac{1}{5}\Gamma_i\Gamma_{kl})
        F_{ab}^{kl} \, \epsilon  + \frac{3m}{20\sqrt{3}} 
        \td\gamma_\ast \td\gamma^{\mu \nu} \gamma^{a}(\Gamma_i{}^j-4\delta_i{}^j)
      \, S_{\mu \nu a ,j} \, \epsilon \, .
\eeq
Using identities \eq{gdiden} and \eq{gddiden} that the gauge-covariantly constant spinor $\eta$ satisfies, we can convert the gamma matrices $\gamma^a$ that are acting on the spinor indices to the gamma matrices $\Gamma^j$ that are acting on the isospin spinor indices. Thus, we get
\beq
\bald
\delta\lambda_i &=
    \frac{1}{32m\lambda^2}  (\Gamma_{kl}\Gamma_i
       -\frac{1}{5}\Gamma_i\Gamma_{kl}) \Gamma^{ab}
        F_{ab}^{kl} \, \epsilon   + \frac{3}{10m^3\lambda^4} \gamma_\ast \gamma^{\mu \nu} \varepsilon_{\mu \nu}  (\Gamma_{ij}-4\delta_{ij})\Gamma^{j}
       \, \epsilon  \, .
\eald
\eeq
The second term vanishes identically since $ (\Gamma_{ij}-4\delta_{ij})\Gamma^{j} = 0$. Commuting the gamma matrices in the first term ($[\Gamma_{kl},\Gamma_i]=-2(\delta_{ik}\Gamma_l-\delta_{il} \Gamma_k)$), we get
\bea
\delta\lambda_i  &=&
    \frac{1}{32m\lambda^2}  \left(
       \frac{4}{5}\Gamma_i\Gamma_{kl}-4 \delta_{ik} \Gamma_l \right) \Gamma^{ab}
        F_{ab}^{kl} \, \epsilon  \\
        &=&
    \frac{1}{32m\lambda^2}  (48\Gamma_i-48 \delta_{ik} \Gamma_k )  \, \epsilon =0
\eea
using eqns \eq{Fident} and \eq{Fident2}.

\subsection{Gravitino variation}

\subsubsection{Internal directions}
The gravitino variation \eq{eq:gravvar} in the Wu directions simplifies to
\beq
\bald
\delta\psi_{a} = D_{a}\epsilon + \frac{1}{4}m\,\hat{\gamma}_{a}\epsilon 
& - \frac{1}{40}({\hat{\gamma}_{a}} {}^{NP} 
- 8\delta_{a} {}^{N}\hat{\gamma}^{P})\Gamma_{AB} G_{NP}^{AB} \, \epsilon  \\
&+\frac{m}{10 \sqrt{3}} \left(\hat{\gamma}_a {}^{NPR}- \frac{9}{2} \delta_a {}^N \hat{\gamma}^{PR} \right) \Gamma^A  S_{NPR,A} \, \epsilon\, .
\eald
\end{equation}
The first term immediately vanishes, since $\eta$ is $SO(3)$-gauge covariantly constant. Decomposing the 7D gamma matrices $\hat{\gamma}$ as in eqn \eq{g7d} and $\epsilon$ as in eqn \eq{spinansatz}, and plugging in $S$ from eqn \eq{Ssol}, we get
\beq
\bald
m^{-1}\delta\psi_{a} =&  \frac{1}{4}\,\td\gamma_\ast\, \gamma_{a}\epsilon 
 - \frac{1}{80}C^2\, \td\gamma_\ast\,(\gamma_{a} {}^{bc} 
- 8\delta_{a} {}^{b} \gamma^{c})\Gamma_{AB} F_{bc}^{AB} \, \epsilon  \\
& \qquad \qquad -\frac{1}{5}C^4 \left(3\gamma_a {}^{b} \td\gamma^{\alp \beta} 
-  \frac{9}{2} \delta_a {}^b \td\gamma^{\alp \beta} \right) \Gamma^A S_{b \alp \beta,A} \, \epsilon \, .
\eald
\end{equation}
The second term, after using $\gamma_a {}^{bc} = \gamma_a \gamma^{bc} - 2 \delta_a {}^{[b} \gamma^{c]}$, becomes
\beq
\bald
- \frac{1}{80}C^2\, \td\gamma_\ast\,(\gamma_{a} \gamma^{bc} 
- 10\delta_{a} {}^{b} \gamma^{c})\Gamma_{AB} F_{bc}^{AB} \, \epsilon & = 
- \frac{1}{80}C^2\,\td\gamma_\ast\, (60 \gamma_{a}- 120 \delta_{a} {}^{b} \gamma_b) \, \epsilon \\
&= \frac{3}{4}C^2 \td\gamma_\ast\, \gamma_a \, \epsilon \, ,
\eald
\eeq
where we used the identities \eq{Fident} and \eq{Fident1}. Similarly, the third term is
\beq
\bald
-\frac{1}{5}C^4 \left(3\gamma_a {}^{b}\,\td\gamma^{\alp \beta} 
- \frac{9}{2} \delta_a {}^b \,\td\gamma^{\alp \beta} \right) \Gamma^A S_{b \alp \beta,A}  \, \epsilon
&= -\frac{1}{5}C^4 \varepsilon_{\alp \beta}\, \td\gamma^{\alp \beta} \left( 3\gamma_a {}^{b}-  
\frac{9}{2} \delta_a {}^b \right) \Gamma^A \delta_{bA} \, \epsilon \\
&= -\frac{2}{5}C^4 \,\td\gamma_\ast \left( 3 \gamma_a {}^b \gamma_b -  \frac{9}{2} \gamma_a \right) \epsilon \\
&= -3C^4 \,\td\gamma_\ast \, \gamma_a  \epsilon  \, ,
\eald
\eeq
where we used eqns \eq{gdiden} and \eq{gamma3} in the second line.

Collecting all the terms, we have
\beq
\delta\psi_{a} = \left( \frac{m}{4} + \frac{3}{4 m \lambda^2} - \frac{3}{m^3 \lambda^4} \right) \td\gamma_\ast 
\,\gamma_a \epsilon \, .
\eeq
Thus, setting the gravitino variation to zero in the Wu directions gives the algebraic equation
\begin{equation}
    1+ 3 C^2-12C^4=0 \, ,\label{quadratic}
\end{equation}
which has the physically-allowed solution
\beq
C^2=\frac{\sqrt{57}+3}{24} \, .
\label{lambdasol}
\eeq
Note that eqn \eq{quadratic} also appeared as a factor arising from the Einstein equations in the internal directions, see eqn \eq{einsint}.

\subsubsection{External directions}
\label{sec:gravext}
The gravitino variation \eq{eq:gravvar} in the AdS$_2$ directions is
\beq
\bald
\delta\psi_{\alpha} = \nabla_{\alpha}\epsilon + \frac{1}{4}m\,\hat{\gamma}_{\alpha}\epsilon 
& - \frac{1}{40}({\hat{\gamma}_{\alpha}} {}^{NP} 
- 8\delta_{\alpha} {}^{N}\hat{\gamma}^{P})\Gamma_{AB} G_{NP}^{AB} \, \epsilon  \\
&+\frac{m}{10 \sqrt{3}} \left(\hat{\gamma}_\alpha {}^{NPQ}- \frac{9}{2} \delta_\alpha {}^N \hat{\gamma}^{PQ} \right) \Gamma^A  S_{NPQ,A} \, \epsilon \, .
\eald
\end{equation}
The first term becomes only a covariant derivative since the gauge potential in our ansatz does not have components in the AdS$_2$ directions. Assuming $\xi$ is a Killing spinor on AdS$_2$, satisfying $\nabla_\alp \xi=c \, \td\gamma_\alp \,\xi$, and decomposing the 7D gamma matrices, we get
\beq
\bald
\delta\psi_{\alpha} =c \, \td\gamma_{\alpha}\,\epsilon + \frac{1}{4}m\,\td\gamma_{\alpha}\,\epsilon 
& - \frac{1}{80m \lambda^2}({\td\gamma_{\alpha}} \gamma^{ab} 
- 8\delta_{\alpha} {}^{a}\gamma^{b})\Gamma_{AB} F_{ab}^{AB} \, \epsilon  \\
&-\frac{1}{5m^3 \lambda^4} \left(3\td\gamma_\alpha {}^{\bet \delta} \,\gamma^a- 9 \delta_\alpha {}^\bet \,\td\gamma^{\delta} \gamma^{a} \right) \td\gamma_\ast  \Gamma^A  \varepsilon_{ \bet \delta} \delta_{aA} \, \epsilon \, .
\eald
\end{equation}
The second term is
\beq
\bald
- \frac{1}{80m \lambda^2}({\td\gamma_{\alpha}} \gamma^{ab} 
- 8\delta_{\alpha} {}^{a}\gamma^{b})\Gamma_{AB} F_{ab}^{AB} \, \epsilon  = 
- \frac{3}{4m \lambda^2}{\td\gamma_{\alpha}}  \, \epsilon 
\eald
\eeq
using \eq{Fident} and $\delta_\alp^a=0$. The last term in the variation is then
\beq
\bald
\frac{9}{5 m^3 \lambda^4}   \delta_\alpha {}^\bet \,\td\gamma^{\delta} \, \varepsilon_{ \bet \delta}  
\,\td\gamma_\ast \gamma^{a}  \Gamma^A  \delta_{aA} \, \epsilon  &=\frac{9}{5 m^3 \lambda^4} \,\td\delta_\alpha {}^\bet \,\td\gamma^{\delta} \, \varepsilon_{ \bet \delta}\, \td\gamma_\ast  \,\gamma^{a}  \gamma^A  \delta_{aA} \, \epsilon \\
&=\frac{9}{m^3 \lambda^4}\, \td\gamma_\alpha \, \epsilon \, ,
\eald
\eeq
where we used eqn \eq{gdiden} to convert $\Gamma \epsilon$ to $\gamma \ep$, and $\td\gamma^{\delta}\,  
\varepsilon_{\alp \delta} \, \td\gamma_\ast = \td\gamma_\alp$. Hence, setting the gravitino variation to zero in the 
AdS$_2$ directions gives
\begin{equation}
\left(\frac{c}{m}+\frac{1}{4}-\frac{3}{4}C^2+9C^4\right)\td\gamma_\alpha \,\epsilon=0 \, ,\label{AdS2sign}
\end{equation}
for an AdS$_2$ Killing spinor $\xi$ obeying
\begin{equation}
    \nabla_\alpha \xi = c \,\td\gamma_\alpha\, \xi =-\frac{\sqrt{-\Lambda}}{2}\, \td\gamma_\alpha \xi \, .
    \label{AdSKil}
\end{equation}
Hence, the cosmological constant for AdS$_2$ can be written as
\beq
\Lambda \lambda^2=-\frac{19}{8}(\sqrt{57} +3) \, ,
\label{Lambdasol}
\eeq
where $\lambda$ may be substituted using eqn~\eq{lambdasol}. Note that this solution preserves 1/16 supersymmetry since there are 2 Killing spinors satisfying eqn \eq{AdSKil} and only one gauge-covariantly constant spinor on the Wu manifold.

\subsection{Non-compact dual of the Wu manifold, \texorpdfstring{$SL(3,\R)/SO(3)$}{SL(3,R)/SO(3)}}

The construction in this section still works if we replace the Wu manifold with its non-compact dual, 
the coset $SL(3,\R)/SO(3)$. We can obtain the metric on $SL(3,\R)/SO(3)$ by redefining the two radial coordinates
$\mu$ and $\nu$ in the Wu metric (see appendix \ref{wurecap}) according to
\beq
\mu\longrightarrow \im \mu\,,\qquad \nu\longrightarrow \im \nu\,, \label{cnc}
\eeq
while sending $ds_\text{Wu}^2\longrightarrow 
d s_{SL(3,\R)/SO(3)}^2= -ds_\text{Wu}^2$. Under these identifications, the only necessary differences are \cite{Wu}
\beq
\bar{R}_{ab} \rightarrow \tilde{R}_{ab}=-6 \delta_{ab} \, , \quad
F^i_{ab} \rightarrow \tilde{F}^i_{ab} = -F^i_{ab} \, ,
\label{duality}
\eeq
where $\tilde{R}_{ab}$ is the Ricci tensor and $\tilde{F}^i$ are the $SO(3)$ gauge fields on $SL(3,\R)/SO(3)$. Note that the equation of the integrability condition for the $SO(3)$-gauge covariantly constant spinor is the same as in the Wu manifold \cite{Wu}. Thus, we can use the same $\epsilon$ with the same properties in the variations of the fermionic fields.

Let us start with the same ansatz, only replacing the metric on the Wu manifold with the one on $SL(3,\R)/SO(3)$:
\beq
ds_7^2 = ds_{\text{AdS}_2}^2(x) + \tilde{\lambda}^2 ds_{SL(3,\R)/SO(3)}^2(y) \, ,
\eeq
and
\begin{equation}
G_{ab}^{AB} = \frac{1}{2m \tilde{\lambda}^2} \tilde F_{ab}^{AB} = -\frac{1}{2m \tilde{\lambda}^2} F_{ab}^{AB}\, .
\label{WuFieldEmbed2}
\end{equation}
We obtain the same solution for $S$ as in eqn \eq{Ssol} when we solve its equation of motion, eqn \eq{Seom}. Thus, defining $\tilde{C}=(m\tilde{\lambda})^{-1}$, the Einstein equations in the internal directions have only a sign change on the left-hand side:
\beq
\bald
-6\delta_{ab} &=  \left(- \frac{3}{2\tilde{C}^2} + \frac{15}{2}\tilde{C}^2 + 72\tilde{C}^4 \right) \delta_{ab}   \, ,
\eald
\eeq
which factorises to
\beq
\left(  1 - 3 \tilde{C}^2 -12\tilde{C}^4 \right) \left( 1 - \tilde{C}^2 + 4\tilde{C}^4 \right) = 0 \, .
\label{einsint2}
\eeq

The gravitino variation in the internal directions gives
\begin{equation}
    1- 3 \tilde{C}^2-12\tilde{C}^4=0 \, ,\label{quadratic2}
\end{equation}
which has the same roots as the first factor of eqn \eq{einsint2}. It admits a real solution for $\wtd C$, with
$\wtd C^2= \fft{\sqrt{57}-3}{24}$.  Thus, the solution with
\beq
m^2\tilde{\lambda}^2=\frac{\sqrt{57} +3}{2} \, ,
\label{lambdasol2}
\eeq
for the parameter $\tilde \lambda$ solves both the Einstein equations and $\delta \psi_a=0$. 

In the external directions, the Einstein equations stay the same:
\beq
\tilde{\Lambda} \bar{\eta}_{\alp \bet} = \left(- \frac{3}{2}m^2 - \frac{15}{2m^2\tilde{\lambda}^4} - \frac{216}{m^6 \tilde{\lambda}^8} \right) \bar{\eta}_{\alp \bet}  \, ,
\eeq
while the gravitino variation gives
\begin{equation}
\left(\tilde{c}+\frac{m}{4}+\frac{3}{4m\tilde{\lambda}^2}+\frac{9}{m^3\tilde{\lambda}^4}\right)\gamma_\alpha \epsilon=0 \, ,
\end{equation}
for an AdS$_2$ Killing spinor $\xi$ obeying
\begin{equation}
    \nabla_\alpha \xi = \tilde{c} \gamma_\alpha\xi =-\frac{\sqrt{\tilde{\Lambda}}}{2} \gamma_\alp  \xi \, .
\end{equation}
Therefore, we get the cosmological constant for the AdS$_2$ space as
\beq
\tilde{\Lambda} \tilde{\lambda}^2=-\frac{19}{8}(\sqrt{57} -3) \, .
\label{Lambdasol2}
\eeq

It should be noted that the subtleties associated with the spin$^h$ structure of the Wu manifold do not apply in the non-compact case.  The manifold $SL(3,\R)/SO(3)$ in this case is topologically $\R^5$ and there are no non-trivial cycles that could give rise to non-trivial holonomies for spinor fields.

\section{11D supergravity solution\label{D11sec}}

\subsection{Uplift from D=7 to D=11}

It was shown in \cite{Nastase:1999cb,Nastase:1999kf} that eleven-dimensional supergravity admits a consistent reduction on 
$S^4$, giving rise to seven-dimensional maximal gauged supergravity. Thus, any solution of the seven-dimensional theory can be lifted to a solution in eleven-dimensional supergravity.  Furthermore, if the solution in seven dimensions is supersymmetric, then the lifted solution in eleven dimensions will also be supersymmetric. 

In particular, the supersymmetric AdS$_2\times W$ and AdS$_2\times SL(3,\R)/SO(3)$ solutions 
\bea
\label{wusol}
ds_7^2 &=& ds_\text{AdS$_2$}^2 + \lambda^2 ds_\text{Wu}^2 \, , \\
ds_7^2 &=& ds_\text{AdS$_2$}^2 + \tilde \lambda^2 ds_{SL(3,\R)/SO(3)}^2 \, ,
\eea
where $ \lambda$, $\tilde{\lambda}$, and the cosmological constants of AdS$_2$ spaces, $\Lambda$ and $\tilde{\Lambda}$ satisfying eqns \eq{lambdasol}, \eq{lambdasol2}, \eq{Lambdasol} and \eq{Lambdasol2}, respectively, can be uplifted to $D=11$ using the results in \cite{Nastase:1999cb,Nastase:1999kf}.  In fact, the explicit uplift formulae that we shall use are given in section 2 of \cite{cvluposatr}. 
The uplifted $D=11$ solution for the Wu manifold case is
\begin{equation}
\begin{split}
&ds_{11}^2=\lambda^2(e^{2A}d\bar{s}^2_{\text{AdS}_2}+ds^2_9) \, ,\\
&ds_9^2=ds_\text{Wu}^2+C^2DY^ADY^A,
\end{split}
\end{equation}
where $Y^A$ are the coordinates parametrising $S^4$ with the constraint $Y^AY^A=1$, and
\begin{equation}
    DY^A=dY^A+2mB^{AB}Y^B \, ,
\end{equation}
and $d\bar{s}^2_{\text{AdS}_2}$ is unit-radius AdS$_2$, and
\begin{equation}
    e^{-A}=\sqrt{-\Lambda\lambda^2}=\frac{1}{C}+12C^3 \, .
\end{equation}
Note that the four-form field strength in $D=11$ can be found in \cite{Nastase:1999cb,cvluposatr}.

\subsection{AdS\texorpdfstring{$_2\times M_9$}{2 x M9} solution}

We can view our solution as AdS$_2 \times M_9$, with $M_9$ the 9-manifold described by a twisted $S^4$ bundle over the Wu manifold. $M_9$ is non-Einstein and non-homogeneous.  As will be discussed in greater detail below, $M_9$, the $S^4$ 
bundle over the Wu manifold, is a spin manifold, even though
the Wu manifold forming the base does not admit a spin structure.  The reason for this is that the non-vanishing
$w_2$ Stiefel-Whitney obstruction of the Wu manifold is cancelled by a $w_2$ obstruction in the $SO(3)$ action of 
the twisted $S^4$ fibration, resulting in a 9-manifold on which ordinary (uncharged) spinors can be defined.

In this basis,
\begin{equation}
B^{AB}=\frac{1}{2m}A^i(T^i)^{AB} \, .
\end{equation}
Therefore, the metric on the 9-manifold is (with $ds^2_\text{Wu}$ normalised such that $R_{ab}=6\delta_{ab}$)
\begin{equation}
\begin{split}
    d\hat{s}_9^2=& \lambda^2 ds_\text{Wu}^2+\frac{1}{m^2}(dY^A+2mB^{AB}Y^B)(dY^A+2mB^{AC}Y^C)\\
    =&\lambda^2ds_\text{Wu}^2+\frac{1}{m^2}\big(dY^A+A^i(T^i)^{AB}Y^B\big)\big(dY^A+A^j(T^j)^{AC}Y^C\big) \, ,
\end{split}
\end{equation}

In the Wu directions, we switch to a vielbein basis  
\begin{equation}
    \hat{e}^a=\lambda e^a \, .
\end{equation}

We define (as in, for example, \cite{embeddingkiller})
\begin{equation}
    K^{i r}\partial_r Y^A=(T^i)^{AB}Y^B \, ,
\end{equation}
where $r=1,2,3,4$ label the intrinsic coordinates on $S^4$.  If we write the 5-vector $Y$, whose components are $Y^A$, as 
\bea
Y= {\cal O}\, Y_0\,,\qquad {\cal O}= e^{\phi\, T^3}\, e^{\theta\, T^2}\, e^{\psi\, T^3}\,,\qquad Y_0=(0,0,0,\sin\chi,\cos\chi)^T\,,
\eea
then the three Killing vectors $K^i= K^{i r}\, \del_r$ turn out to be given by
\bea
K^1 &=& \sin\phi\, \del_\theta - \cos\phi\, \cot\theta\, \del_\phi + \cos\phi\,\csc\theta\, \del_\psi\,,\nn\\
K^2 &=& \cos\phi\, \del_\theta - \sin\phi\, \cot\theta\, \del_\phi + \sin\phi\, \csc\theta\, \del_\psi\,,\nn\\
K^3 &=& \del_\phi\,.
\eea
In terms of the coordinates $(\theta,\phi,\psi,\chi)$ introduced above, the metric $d\Omega_4^2 = dY^A\, dY^A$ on the unit
4-sphere takes the triaxial Bianchi IX form
\bea
d\Omega_4^2 = d\chi^2 + 4\sin^2(\chi+ \ft{2\pi}{3})\, \Sigma_1^2 + 4 \sin^2(\chi-\ft{2\pi}{3})\, \Sigma_2^2 
+ 4\sin^2\chi\, \Sigma_3^2\,,
\eea
where the $\Sigma_i$ are the left-invariant 1-forms of $SO(3)$, defined by
\bea
{\cal O}^{-1} d{\cal O} = T^i\, \Sigma_i\,,
\eea
implying
\bea
\Sigma_1 &=&\sin\psi\, d\theta - \cos\psi\, \sin\theta\, d\phi\,,\nn\\
\Sigma_2 &=& \cos\psi\, d\theta + \sin\psi\, \sin\theta\, d\phi\,,\nn\\
\Sigma_3 &=& d\psi + \sin\theta\, d\phi\,.
\eea
In the complete ``unit'' metric $ds_4^2= (dY^A + A^{AB}\, Y^B)^2$ in the fibre directions, it can be seen from the definitions above that
\bea
ds_4^2 &=& d\chi^2 + 4 \sin^2(\chi + \ft{2\pi}{3})\, (\Sigma_1 +\wtd A^1)^2 + 4\sin^2(\chi-\ft{2\pi}{3})\, (\Sigma_2+\wtd A^2)^2 \nn\\
 && +4\sin^2\chi\, (\Sigma_3+ \wtd A^3)^2\,,
\eea
where $\wtd A^i$ is related to $A^i$ by
\bea
\wtd A^i\, T^i = {\cal O}^{-1}\, A^i\, T^i\, {\cal O}\,.
\eea

We then switch to a vielbein basis via
\begin{equation}
    K^\alpha_i=\hat{e}^\alpha_r K^r_i \, .
\end{equation}

Then we can write components of the Ricci tensor using the standard Kaluza-Klein decomposition
\begin{equation}
    \begin{split}
        &\hat{R}_{ab}=R_{ab}-\frac{1}{2}K^{\alpha i}K_\alpha^j\mathcal{F}_{a c}^i\mathcal{F}_b ^{jc} \, ,\\
        &\hat{R}_{a \alpha}=\frac{1}{2}K_\alpha^iD_b \mathcal{F}_a^{i b} \,,\\
        &\hat{R}_{\alpha \beta}=R_{\alpha \beta}+\frac{1}{4}K_\alpha^iK_\beta^j\mathcal{F}_{ab }^i\mathcal{F}^{jab} \, .
    \end{split}
\end{equation}
An explicit calculation of the Ricci scalar is given in appendix~\ref{kkcalc}, with a fuller list of curvature invariants given later in eqn~(\ref{ci}).

\subsection{Cohomogeneity-one description of 9-manifold}

As in \cite{cosets}, we can consider the following. We identify the Lie algebra of $SU(3)$, denoted by $\mathfrak{su}(3)$, with
traceless Hermitean $3\times 3$ matrices $T^A{}_B$, and hence
the left-invariant 1-forms $L_A{}^B$ are complex valued, with
$L_A{}^A=0$ and $(L_A{}^B)^\dagger = L_B{}^A$.  They satisfy the algebra
\be
dL_A{}^B = \im\, L_A{}^C\wedge L_C{}^B\,.
\ee

   It is convenient to decompose the $\mathfrak{su}(3)$ algebra with respect to
its maximal $\mathfrak{so}(3)$ subalgebra:
\bea
K_1 &=& L_2{}^3 + L_3{}^2\,,\qquad
K_2= L_3{}^1 + L_1{}^3\,,\qquad
K_3= L_1{}^2 + L_2{}^1\,,\nn\\
K_4 &=& L_1{}^1 - L_2{}^2\,,\qquad
K_5= \fft1{\sqrt3}\, (L_1{}^1 + L_2{}^2 - 2 L_3{}^3)\,,\nn\\
H_1 &=& \im\, (L_3{}^2 - L_2{}^3)\,,\qquad
H_2= \im\, (L_1{}^3 - L_3{}^1)\,,\qquad
H_3= \im\, (L_1{}^2 - L_2{}^1)\,.
\eea
(We have reversed the signs of $H_1$ and $H_2$ from those given in \cite{cosets}, in order to fit more conveniently
with our present conventions.)
The subalgebra $\mathfrak{so}(3)_{\rm maximal}$ 
is generated by $H_1$, $H_2$ and $H_3$, and the $K_i$ transform as a
$\bf 5$ under $\mathfrak{so}(3)_{\rm maximal}$.  Note that the 1-forms $K_i$ and $H_i$ obey the algebra
\bea
dH_i &=& -\ft14 \epsilon_{ijk}\, H_j\wedge H_k + \ft14 T^i_{ab}\, K_a\wedge K_b\,,\nn\\
dK_a &=& -\ft12 T^i_{ab}\, H_i\wedge K_b\,,
\eea
where $T^i_{ab}$ are the usual generators of the $SO(3)$ subgroup of $SO(5)$, in the $5\times 5$ representation, as in \cite{Wu}. Thus the
non-zero components are specified by
\bea
T^1_{14}=-T^1_{23}=1\,,\ T^1_{15}=-\sqrt3\,; \ T^2_{24}=T^2_{13}=1\,,\  T^2_{25}=\sqrt3\,; \
T^3_{12}=1\,, \ T^3_{34}=2\,, 
\eea
together with those consequent from antisymmetry.

Following \cite{gonzalez}, $M_9$, the twisted $S^4$ bundle over the Wu manifold, can be described as a cohomogeneity-one manifold, by writing
\begin{equation}
    ds_{11}^2=ds_{\text{AdS}_2}^2+\lambda^2ds_9^2,
\end{equation}
where
\bea
    &&ds^2_9=\frac{1}{4}(K_1^2+K_2^2+K_3^2+K_4^2+K_5^2)\nn\\
&&\qquad + C^2\, \Big[d\chi^2 + \sin^2\big(\chi +\fft{2\pi}{3}\big)\, H_1^2 +
 \sin^2\big(\chi -\fft{2\pi}{3}\big)\, H_2^2 + \sin^2\chi\, H_3^2\Big]\,,\label{cohom1metric}
\eea
where $C$ was defined in eqn (\ref{Cdef}), i.e. $C=(m \lambda)^{-1}$.
The level surfaces at generic constant values of $\chi$ are the homogeneous 8-dimensional space $SU(3)/(\Z_2\times \Z_2)$, 
where the $SU(3)$ is quotiented by the discrete diagonal $SU(3)$ matrices diag$(\epsilon_1,\epsilon_2,\epsilon_3)$ where 
$\epsilon_i=\pm1$ and $\epsilon_1\epsilon_2\epsilon_3=1$ \cite{gonzalez}. The
coordinate $\chi$ ranges between two adjacent endpoints where one or another of the three $\chi$-dependent functions in
eqn (\ref{cohom1metric}) goes to zero.  We can choose, for example, $0\le \chi\le \ft13\pi$. As the singular orbits
at $\chi=0$ and $\chi=\ft13\pi$ are approached, the metric in the neighbourhood extends smoothly onto $\R^2\times N(1,-1)/\mathbb{Z}_2$, where $N(1,-1)$ is the Aloff-Wallach space $SU(3)/U(1)_{(1,-1)}$. (The Aloff-Wallach space $N(k,\ell)=SU(3)/U(1)_{(k,\ell)}$ is
the quotient of $SU(3)$ by the $U(1)$ subgroup comprising matrices of the form diag$(e^{\im k\theta}, e^{\im \ell\theta},
e^{-\im(k+\ell)\theta})$, where $k$ and $\ell$ are relatively-prime integers. \cite{Aloff})

A calculation of curvature invariants yields
\bea
\hat{R}&=&m^2 (12+ 30 C^2 -15 C^4)\,, \nn\\
\hat{R}_{ab}\hat{R}^{ab}&=&18 m^4 \left(2+ 15 C^4 -20 C^6 +18 C^8\right)\,,\nn\\
\hat{R}_{abcd}\hat{R}^{abcd}&=& m^4 \left(24 +840 C^4 -990 C^6 +\frac{2385}{2} C^8\right)\,,\nn\\
\hat{R}^a_b\hat{R}_a^c\hat{R}^b_c&=&\frac{3m^6}{2}(72 + 270 C^4 + 720 C^6 -1485 C^8+2538 C^{10} \nn\\
&&\qquad -473 C^{12} -49 C^{12}\, \cos6\chi). \label{ci}
\eea
It is evident from the fact that the invariant $\hat{R}^a_b\hat{R}_a^c\hat{R}^b_c$ is not constant, but rather, that
it depends on the coordinate $\chi$ which parameterises the homogeneous 8-dimensional level surfaces, that the metric
$ds_9^2$ on the $S^4$ bundle over the Wu manifold is inhomogeneous.  (In other words, the metric is of strict cohomogeneity one.)

\subsection{\texorpdfstring{$SU(4)$}{SU(4)} structure}

Here, we show that our eleven-dimensional supergravity solution fits into a class of AdS$_2$ solutions which admit an $SU(4)$ structure that was described in \cite{zayas}, which was built on the classification of 11D Killing spinors in \cite{11dspinor1,11dspinor2} using a derivation similar to \cite{bubbling,Katmadas:2015ima}. To do this, we choose a numbering for the 
vielbein basis $e^{\td a}$ for the nine-dimensional metric in eqn (\ref{cohom1metric}) 
that matches the precise conventions in \cite{zayas}:
\bea
e^1 &=& C\sin\left(\chi+\frac{2\pi}{3}\right)H_1\,, \enspace e^2= \ft12 K_1\,,  
\enspace e^3= C\sin\left(\chi-\frac{2\pi}{3}\right)H_2\,, \enspace e^4=\ft12 K_2\,,\nn\\
\enspace e^5 &=& C\, \sin\chi\,H_3\,, \enspace e^6= -\ft12 K_3\,,
\enspace e^7 =\ft12(\cos \chi\,K_4- \sin \chi\,K_5)\,, \enspace e^8=C\, d\chi\,, \nn\\
e^9 &=& \ft12(\sin \chi\, K_4+ \cos \chi\, K_5)\,.
\eea

Then the solution can be written as
\begin{equation}
\begin{split}
&ds_{11}^2=\lambda^2(e^{2A}d\bar{s}^2_{\text{AdS}_2}+ds_9^2) \, ,\\
&F_4= \lambda^{3}\, (G_4+e^{2A}\overbar{\text{Vol}}_{\text{AdS}_2}\wedge G_2) \, ,\\
&e^{-A}=\sqrt{-\Lambda \lambda^2}=\frac{1}{C}+12C^3 \, ,
\end{split}
\end{equation}
where $d\bar{s}^2_{\text{AdS}_2}$ is the metric of unit-radius AdS$_2$ with volume form $\overbar{\text{Vol}}_{\text{AdS}_2}$, and where $G_2$ and $G_4$ are closed 2- and 4-forms ($dG_2=0$ and $dG_4=0$) satisfying
\beq
\bald
d \ast G_{4} + G_{2} \wedge G_{4} = 0  \, ,\\
d \ast G_{2} - \frac{1}{2} G_{4} \wedge G_{4} =0 \,.\label{G2G4eom}
\eald
\eeq
We can then write the 9-metric as
\begin{equation}
\begin{split}
    &ds_9^2=ds_8^2+V^2,\\
    &ds_8^2=\sum_{i=1}^8(e^i)^2, \quad V=e^9.
    \end{split}
\end{equation}

As in \cite{zayas}, we have
\begin{equation}
    \begin{split}
        &V=e^9,\\
        &J=e^1\wedge e^2 + e^3 \wedge e^4 + e^5 \wedge e^6+e^7\wedge e^8,\\
        &\Omega=(e^1+\im e^2)\wedge (e^3+\im e^4)\wedge (e^5+\im e^6)\wedge (e^7+\im e^8).
    \end{split}
\end{equation}
We have $dJ=0$, and $J$ gives an almost complex structure $J_a{}^b$ on the metric 
$ds_8^2$.\footnote{It should be noted that $ds^2_8$ is a singular metric; the vector field $V^\#$ dual to the 1-form $V$ generates orbits in the maximal 2-torus in $SU(3)$ that depend upon $\chi$ and that are not, in general, even closed.} 
We also have 
\begin{equation}
    \begin{aligned}
        &G_2=-12C^3J=12C^4dV,\\
        &G_4=\frac{C^3}{8}\epsilon_{ABCDE} \,Y_0^A Q^{BC} \wedge Q^{DE},\\
        &Q^{AB}=D_H Y_0^A\wedge D_HY_0^B+\frac{1}{8}F^i_{ab}F^{iAB} \, K^a\wedge K^b,\\
        &D_HY_0^A=dY_0^A + \frac{1}{2}H_i  F^{iA}{}_B \,Y_0^B, \quad Y_0=(0,0,0,\sin \chi, \cos \chi)^T.\label{G2etc}
    \end{aligned}
\end{equation}
With these definitions, the conditions in \cite{zayas} are satisfied, stated in appendix~\ref{pz} for completeness, when $C=(m\lambda)^{-1}$ satisfies
\beq
-12C^4+3C^2+1=0 \, ,\label{Cpoly}
\eeq
which is consistent with the supersymmetric solution in eqn \eq{lambdasol}.
Together with the holomorphic 4-form $\Omega$, for which $\Omega\wedge\bar\Omega=16 \text{Vol}_8$,
this shows that $M_9$ has an $SU(4)$ structure. 

$G_4$ can be written explicitly as
\beq
\begin{aligned}
G_{4}=
& -12 \,C^3 \,\, e^{2}\wedge e^{4}\wedge e^{6}\wedge e^7 + \frac{3}{C} \,e^{1}\wedge e^{3}\wedge e^{5}\wedge e^{8} \\
& \,\quad+C \Big[
 F^1\wedge \Big(e^{1}\wedge e^{8} +2\cos\big(\chi+\tfrac{2\pi}{3}\big)\,e^{3}\wedge e^{5}\Big) \\
& \qquad + F^2\wedge\Big(e^3\wedge e^8 -2\cos\big(\chi-\tfrac{2\pi}{3}\big)\,e^{1}\wedge e^{5}\Big)
+ F^3\wedge\!\Big(e^{5}\wedge e^{8} + 2\cos\chi\;e^{1}\wedge e^{3}\Big)
\Big]\,,\label{G4res}
\end{aligned}
\eeq
where $F^i=\ft18 F^i_{ab}\, K^a\wedge K^b$, i.e.
\beq
\begin{aligned}
F^1 &= \frac{1}{4}( K^{1}\wedge K^{4}-K^{2}\wedge K^{3}  - \sqrt{3}\, K^{1}\wedge K^{5} )\, ,\\
F^2 &= \frac{1}{4}(K^{1}\wedge K^{3} + K^{2}\wedge K^{4} + \sqrt{3}\, K^{2}\wedge K^{5} )\, ,\\
F^3 &= \frac{1}{4}(K^{1}\wedge K^{2} + 2\, K^{3}\wedge K^{4}) \, .
\end{aligned}
\eeq

As will be discussed below in section \ref{topsec}, the 9-manifold $M_9$ has non-vanishing real cohomology $H^n(M_9;\R)$ for
$n=0,4,5, 9$.  The 4-form $G_4$ gives a representation of the generator of $H^4(M_9;\R)=\R$, as can be seen from the fact
that the second term in the expression for $G_4$ in eqn (\ref{G4res}) is proportional to the volume form on the $S^4$ fibres, 
and so the integral of $G_4$ over the homology 4-cycle that generates $H_4(M_9;\R)$ is non-zero.  Although $G_4$ is closed, $dG_4=0$, it is not co-closed. By the Hodge theorem, the harmonic representative $\omega_4$ of $H^4(M_9;\R)$ will therefore have the form 
$\omega_4= G_4 + d\beta$, where $\beta$ is some globally-defined 3-form.  It turns out that $\beta$ can be written
as a certain constant multiple of $\ast (J\wedge G_4)$. Specifically, if we define
\begin{equation}
    \omega_4\equiv G_4-\left(\frac{C^5}{1+C^4}\right)d \ast(J \wedge G_4),
\end{equation}
then $\omega_4$ satisfies
\begin{equation}
    d\omega_4=d\ast \omega_4=0.
\end{equation}

\subsection{Supersymmetry in $D=11$}

The supersymmetry of the lifted solution can be verified rather straightforwardly in eleven dimensions.
The gravitino transformation rule in $D=11$ takes the form
\bea
\delta\hat\psi_{\hat A} = \hat\nabla_{\hat A}\, \hat\epsilon + \fft1{24} \big(3 \hat F\, \hat\Gamma_{\hat A} - 
\hat\Gamma_{\hat A}\, \hat F\big)\,
\hat\epsilon\,,\qquad \hbox{where}\quad \hat F=\fft1{4!}\, \hat F_{\hat A\hat B\hat C\hat D}\, 
\hat\Gamma^{\hat A\hat B\hat C\hat D}\,,\label{D11susy}
\eea
where we are putting hats on the eleven-dimensional quantities. We decompose the Dirac matrices as the product of the $2\times2$ matrices $\td\gamma_\alpha$ for AdS$_2$ and $16\times16$ matrices $\wtd\Gamma_{\td a}$ for $M_9$ as
\bea
\hat\Gamma_\alpha= \td\gamma_\alpha\otimes \oneone\,,\qquad \hat\Gamma_{\td a}= \td\gamma_\ast\otimes \wtd\Gamma_{\td a}\,,
\eea
with the indices being decomposed as $\hat A=(\alpha,\td a)$ on AdS$_2$ and $M_9$ respectively.
For the Dirac matrices on $M_9$ we shall use the basis
\bea
\wtd\Gamma_{\td a}= \begin{cases} \gamma_{\td a}\times \Gamma_5 &\text{for}\quad 1\le \td a\le 5\\
                                        \oneone\times \Gamma_{\td a-5} & \text{for} \quad 6\le \td a\le 9\,,
                                        \end{cases}
\eea
where the $4\times 4$ Dirac matrices $\gamma_a$ and $\Gamma_i$ are defined in eqns (\ref{gammadef}) and (\ref{gaugegamma}).

Killing spinors $\hat\epsilon$ in the eleven-dimensional supergravity can then be constructed as follows.  First, 
we note that the matrix $\hat F$ in eqn (\ref{D11susy}) takes the form
\bea
\hat F = \td\gamma_\ast\otimes \boldsymbol{G}_2 + \oneone\otimes \boldsymbol{G}_4\,,\label{FhatGG}
\eea
where $\boldsymbol{G}_2$ and $\boldsymbol{G}_4$ are defined in terms of the 2-form and 4-form fields $G_2$ and $G_4$ by
\bea
\boldsymbol{G}_2 \equiv \ft12 G_{ab}\, \wtd\Gamma^{ab}\,,\qquad 
\boldsymbol{G}_4 \equiv \ft1{24} G_{abcd}\,\wtd\Gamma^{abcd}\,.
\eea

In the nine internal directions, the supersymmetry transformations are
\bea
\delta\hat\psi_{\td a} = \hat\nabla_{\td a}\hat\epsilon +
\ft{1}{24} \Big[3\, \oneone\otimes \boldsymbol{G}_2\,\wtd\Gamma_{\td a} 
- \oneone\otimes\wtd\Gamma_{\td a}\,\boldsymbol{G}_2 +
      3\td\gamma_\ast\otimes \boldsymbol{G}_4\,\wtd\Gamma_{\td a} -
       \td\gamma_\ast\otimes \wtd\Gamma_{\td a}\,\boldsymbol{G}_4\Big]\,\hat\epsilon\,.\label{intsusy}
\eea
In the AdS$_2$ directions, we have
\bea
\delta\hat\psi_\alpha = \hat\nabla_\alpha\,\hat\ep +
    \ft1{12}\Big[\td\alpha\otimes\boldsymbol{G}_4 -2 \td\gamma_\alpha\, \td\gamma_\ast\otimes\boldsymbol{G}_2\Big]\,\hat\ep\,.
\label{extsusy}
\eea
Let $\zeta_\pm$ be chiral spinors in AdS$_2$, obeying the coupled Killing-spinor equations 
\bea
\nabla_\alpha\,\zeta_\pm= c\, \td\gamma_\alpha\, \zeta_\mp\,,\qquad \td\gamma_\ast\,\zeta_\pm= \pm\zeta_\pm\,.\label{AdS2ks}
\eea
(The linear combinations $\xi_\pm = (\zeta_+ \pm \zeta_-)$, which are not chiral eigenstates, 
then obey the standard Killing spinor equations
$\nabla_\alpha\, \xi_\pm= \pm c\, \gamma_\alpha\, \xi_\pm$.)  Note that there are two independent Killing spinors
$\xi_+$ and also two independent $\xi_-$.

Substituting the expressions for the 2-form $G_2$ given in eqn (\ref{G2etc}), and the 4-form $G_4$ given in eqn (\ref{G4res}),
we then find from the internal components (\ref{intsusy}) of the supersymmetry variations that for $\delta\hat\psi_{\td a}=0$
the parameter $\hat\epsilon$ must take the form
\bea
\hat\epsilon = \beta_1\, \zeta_+\otimes \eta_+ + \beta_2\,\zeta_-\otimes\eta_-\,,\label{epsum}
\eea
where $\beta_i$ are constants and
\bea
\eta_+ &=& (0,0,1,0,0,-1,0,0,-1,0,0,0,0,0,0,1)^T\,,\nn\\
\eta_- &=& (0,0,1,0,0,-1,0,0,1,0,0,0,0,0,0,-1)^T\,.\label{etapm}
\eea
(There is also an implicit summation in eqn (\ref{epsum}) 
over the two independent AdS$_2$ Killing spinors.)  The spinors $\eta_\pm$ obey
\bea
\boldsymbol{G}_2\, \eta_\pm = \mp 48 C^3\, \eta_\mp\,,\qquad
\boldsymbol{G}_4\, \eta_\pm = \fft{3}{C}\, (1-3C^2 + 4 C^4)\, \eta_\mp\,.
\eea
Note that the internal-space supersymmetry conditions $\delta\hat\psi_{\td a}=0$ also require that $C$ satisfy the usual condition (\ref{Cpoly}).  

From the supersymmetry condition $\delta\hat\psi_\alpha=0$ in the AdS$_2$ directions, given in eqn (\ref{extsusy}),
we find that the spinor in 
(\ref{epsum}) will satisfy this provided that the constants $\beta_1$ and $\beta_2$ satisfy
\bea
c\,\lambda \,\beta_1 + \fft1{4C}\, (1-3C^2 + 36 C^4)\, \beta_2=0\,,\qquad
c\,\lambda\, \beta_2 + \fft1{4C}\, (1-3C^2 + 36 C^4)\, \beta_1=0\,.\label{be1be2eqns}
\eea
In view of eqn (\ref{AdS2sign}), we have
\bea
c\,\lambda = - \fft1{4C}\, (1-3C^2 + 36 C^4)\,,
\eea
and so eqns (\ref{be1be2eqns}) do indeed have a solution, with $\beta_2=\beta_1$.  Thus from eqn (\ref{epsum}), we have 
Killing spinors
\bea
\hat\epsilon = \zeta_+\otimes \eta_+ + \zeta_-\otimes\eta_-\,.
\label{kspinwu}
\eea
There are two such spinors, since there are two independent Killing spinors on AdS$_2$.
We assumed that the spinors have constant components ($d \hat \epsilon =0$) while getting \eq{kspinwu}, but by further checking the integrability condition $[\hat D_{\hat A},\hat D_{\hat B}]\,\hat\epsilon=0$ for the operator $\hat{D}_{\hat{A}} \equiv \hat\nabla_{\hat A} + \fft1{24} \big(3 \hat F\, \hat\Gamma_{\hat A} - 
\hat\Gamma_{\hat A}\, \hat F\big)$, we found that these are the only spinors. 
Thus the compactification
of $D=11$ supergravity to AdS$_2\times M_9$, where $M_9$ is the $S^4$ bundle over the Wu manifold, preserves
a fraction $\fft1{16}$ of the original 32 supersymmetries.
One final remark is that in the $S^4$ bundle over $S^5$ case, we find that the solution preserves $\fft1{8}$ supersymmetry as derived in appendix~\ref{S4overS^5}. 
While both cases support an $SU(4)$-structure, the $S^4$ bundle over $S^5$ satisfies further conditions (see section 2.4 of \cite{zayas}) and has a warped K\"ahler-structure, and thus, enhanced supersymmetry.

\subsection{Topology of the 9-Manifold \texorpdfstring{$M_9$}{M9}}
The 9-manifold $M_9$ we are considering is the unit tangent bundle of the Wu manifold $W$,
$M_9 \cong ST(W)$:
\begin{equation}
   S^4 \hookrightarrow ST(W)\overset{\pi}{\longrightarrow}W.
\end{equation}
Splitting into horizontal (base) and vertical (fibre) spaces gives
\begin{equation}
    TM_9 \cong H\oplus V, \quad H \cong \pi^\ast TW.
\end{equation}

\subsubsection{Topology of the Wu manifold}
For any closed manifold $M$ of dimension $m$, its Stiefel-Whitney 
classes are determined by its Wu classes $v_i \in H^i(M;\Z_2)$, which are defined by the property 
\bea
\langle v_i\smile x,[M]\rangle = \langle Sq^i\, x, [M]\rangle\qquad \hbox{for all}\quad x\in H^{m-i}(M;\Z_2)\,.
\label{Wuclassdef}
\eea
Here, $[M]$ denotes the fundamental class of the $m$-manifold: $[M] \in H_m(M;\Z_2)$, and
$Sq^i$ denotes the Steenrod square operator \cite{MilnorStasheff1974,SteenrodEpstein1962}. 
The operator $Sq^i$ takes a mod-2 cohomology class and produces a new one of higher degree:
\bea
Sq^i:\qquad H^n(M; \Z_2) \longrightarrow H^{n+i}(M;\Z_2)\,.
\eea
(The set of operators $Sq^i$ for $0\le i\le n$ can be thought of as interpolating between the identity operator $Sq^0$ that
leaves an $x\in H^n(M;\Z_2)$ unchanged, and $Sq^n$, which yields $Sq^n x = x\smile x = x^2$ in $H^{2n}(M;\Z_2)$.)
The angle brackets in eqn (\ref{Wuclassdef}) denote the Kronecker pairing:
\bea
\langle \text{--}, \text{--}\rangle:\qquad H^m(M;\Z_2)\times H_m(M;\Z_2) \longrightarrow \Z_2\,.
\eea
The Wu formula then gives the Stiefel-Whitney classes in terms of the Wu classes:
\bea
w = Sq(v)\,,\quad \hbox{\ie}\qquad w_k= \sum_{i+j=k} Sq^i(v_j)\,,\label{Wuformula}
\eea
where $Sq$ denotes the total Steenrod square $\sum_i Sq^i$. 

 As is noted in Remark 4.40 of \cite{debray2}, the calculation of $H^\ast(W;\mathbb{Z}/2)$ and the Stiefel-Whitney classes is somewhat scattered in the literature. Here, we state the relevant results and refer to \cite{debray2} for more complete proofs in Lemma 4.41 through to Proposition 4.45, as well as a more complete history of results \cite{Bar65,notspinc,Flo73,Chen:2017}.

 In particular,
\begin{equation}
    H^\ast(W,\mathbb{Z}_2)=\Z_2[z_2,z_3]/(z_2^2,z_3^2),
\end{equation}
which just means that the mod 2 cohomology of Wu is generated by the (unique) generators $z_2$ and $z_3$ of $H^2(W;\Z_2)$ and
$H^3(W;\Z_2)$ respectively; they obey the relations $z_2^2=0$ and $z_3^2=0$. Since $z_2z_3\in H^5(W,\Z_2)$, we have
\begin{equation}
    H^i(W,\Z_2)=(\Z_2,0,\Z_2,\Z_2,0,\Z_2).
\end{equation}

The Wu classes $v_i$ of an $m$-manifold vanish for $i>\ft12 m$, and so on the five-dimensional Wu manifold $W$ only $v_0$, $v_1$ and $v_2$ can be non-zero. Since $H^1(W,\Z_2)=0$, we have $w_1=0$ and hence $v_1=0$. (The vanishing of $w_1$ can also be seen from the fact that $W$ is orientable.)  Using $v_1=0$, eqn (\ref{Wuformula}) gives $w_2=v_2+v_1^2=v_2$. We also have 
$w_4=0$, since $H^4(W,\Z_2)=0$. Using Wu's formula,
\begin{equation}
    w_3=Sq^1(w_2)=z_3.
\end{equation}
Using the Wu relations between Stiefel-Whitney classes, 
\bea
Sq^i(w_j) = \sum_{t=0}^i \binom{j+t-i-1}{t}\, w_{i-t}\, w_{j+t}\,,
\eea
and the fact that $w_4=0$, we get
\begin{equation}
    0=Sq^1(w_4)=w_1w_4+3w_0w_5=3w_5\,,
\end{equation}
and since the coefficient of $w_5$ is odd, it follows that $w_5=0$.
Therefore, the total Stiefel-Whitney class of $W$ is
\begin{equation}
    w(W)=1+z_2+z_3.
\end{equation}

The integer cohomology of Wu is given by (see, for example, appendix A of \cite{Lee:2020ojw})
\begin{equation}
    H^i(W,\Z)=(\Z,0,0,\Z_2,0,\Z), \quad \quad i=0,1,2,3,4,5.
\end{equation}
The real cohomology of Wu is then given by
\begin{equation}
    H^i(W,\R)=(\R,0,0,0,0,\R), \quad \quad i=0,1,2,3,4,5\,,
\end{equation}
since tensoring the integer cohomology with $\R$ kills the torsion.

\subsubsection{Stiefel-Whitney classes of $M_9$}

Adding a trivial line bundle restores the unit tangent bundle to the full tangent bundle
\begin{equation}
    V \oplus \R=\pi^\ast TW.
\end{equation}
We then have
\begin{equation}
    TM_9 \oplus \R \cong \pi^\ast TW \oplus \pi^\ast TW     \label{10res} 
\end{equation}

Adding a trivial line bundle does not change the Stiefel-Whitney classes due to the Whitney sum formula. We can then use the Whitney sum formula again, as well as $z_2^2=0$ and $z_3^2=0$ to get
\begin{equation}
\begin{split}
w(M_9)
    =&w(TM_9\oplus \R)\\
    =&\pi^*(w(TW)^2)\\
    =&\pi^*((1+z_2+z_3)^2)\\
    =&\pi^*(1+z_2^2+z_3^2)\\
    =&1.
    \end{split}
\end{equation}
Note that, when expanding the brackets, the cross-terms $2z_2$, $2z_3$ and $2z_2z_3$ all vanish mod 2, implying 
that all the positive Stiefel-Whitney classes vanish, $w(M_9)=1$. In particular, $w_1(M_9)=0$ and $w_2(M_9)=0$ together imply that $M_9$ is a spin manifold.

\subsubsection{Cohomology of $M_9$}
\label{topsec}

As in Property 9.4. of \cite{MilnorStasheff1974}, since $TW$ has odd rank, twice its Euler class $e(TW)\in H^5(W,\Z)$ vanishes or $2e(TW)=0$. Since $H^5(W,\Z)=\Z$, it follows that $e(TW)=0$.

The standard way to calculate the cohomology of the unit tangent bundle of a manifold is to use a Gysin long exact sequence \cite{Hatcher2002,BottTu1982}.
For our case, this takes the form
\bea
\cdots \longrightarrow H^{n-5}(W) \overset{\smile \,e}{\longrightarrow} H^n(W) \overset{\pi^\ast}{\longrightarrow}
H^n(ST(W)) \overset{\pi_\ast}{\longrightarrow} H^{n-4}(W) \overset{\smile \,e}{\longrightarrow} 
H^{n+1}(W)\longrightarrow\cdots\,,\nn\\
\label{Gysin}
\eea
where $\smile \,e$ is the cup product with the Euler class of $W$. 
In particular, since the Euler class vanishes, the Gysin sequence splits into short exact sequences and we get (as in \cite{split})
\begin{equation}
    0\longrightarrow H^i(W)\overset{\pi^\ast}{\longrightarrow} H^i(M_9)\overset{\pi_\ast}\longrightarrow H^{i-4}(W)\longrightarrow 0.
\end{equation}

Since for all $i$, at least one of $H^i(W)$ or $H^{i-4}(W)$ is zero, we get
\begin{equation}
    H^{i}(M_9)\cong H^i(W)\oplus H^{i-4}(W).
\end{equation}

Noting that this holds for all of $\Z,\Z_2,\R$ cohomologies (as in \cite{reiser}, Lemma B.3.1.), we can then write down the cohomology of $M_9$ as in table \ref{tab:2}.

\begin{table}[h!]
\centering
\begin{tabular}{c|c|c|c|c|c|c|c|c|c|c}  
\toprule
$n=$ & 0 & 1& 2 &3 &4 &5 &6 &7 &8 &9\\
\hline
$H^n(M_9;\Z_2)=$& $\Z_2$ & 0 & $\Z_2$ & $\Z_2$ & $\Z_2$ & $\Z_2$ & $\Z_2$ & $\Z_2$ & 0 &$\Z_2$\\
\hline
$H^n(M_9;\Z)=$& $\Z$ & 0 & 0 & $\Z_2$ & $\Z$ & $\Z$ & 0 & $\Z_2$ & 0 &$\Z$\\
\hline
$H^n(M_9;\R)=$& $\R$ & 0 & 0 & 0 & $\R$ & $\R$ & 0 & 0& 0 &$\R$\\
\bottomrule
\end{tabular}
\caption{\label{tab:2}Cohomology groups of the 9-manifold $M^9$.}
\end{table}

These satisfy the usual relations
\begin{equation}
\begin{split}
&\hbox{Free}\big(H^i(M_9;\Z)\big) \cong \hbox{Free}\big(H^{9-i}(M_9,\Z)\big),\\
& \hbox{Tor}\big(H^i(M_9;\Z)\big) \cong \hbox{Tor}\big(H^{10-i}(M_9,\Z)\big)\,.
\end{split}
\end{equation}

It is interesting to note that analogous calculations for the unit tangent bundle over $S^5$ lead to 
\begin{equation}
    H^n(ST(S^5);F)=(F,0,0,0,F,F,0,0,0,F),
\end{equation}
with the field $F$ being any of $\Z_2$, $\Z$ or $\R$. This would be the resulting manifold from uplifting the AdS$_2\times S^5$ solution of 7D gauged supergravity in \cite{Gauntlett:2000ng} to 11D. In this case, $ST(S^5)$ is the homogeneous Stiefel manifold $V_2(\R^6)=SO(6)/SO(4)$ which is a $U(1)$ bundle over the K\"ahler manifold $SO(6)/(SO(2)\times SO(4))$. Note that in this $S^5$ case, unlike the lifting of the Wu solution that we have been discussing previously, the 9-manifold is a regular $U(1)$ bundle over an 8-manifold that is actually smooth.  We discuss some further details of this $S^5$ example in appendix 
\ref{S4overS^5}.

\subsection{Flux quantisation}

For $M_9$ in the 11D solution involving the Wu manifold, the 4-form $F_4$ contains a term proportional to $G_4$ in $H^4(M_9,\R)$ that can be paired with the $S^4$ fibre in $H_4(M_9,\R)$ to give an M5 brane conserved magnetic charge. The term proportional to $\overbar{\text{Vol}}_{\text{AdS}_2}\wedge G_2$ cannot result in an M2 brane conserved electric charge since the dual of $G_2$ (actually with another term to form the Page charge) would need to be integrated over a 7-cycle but $H_7(M_9,\R)\cong H^2(M_9,\R)=0$, so this cannot lead to a charge.
\section{Generalised BPS flow}
\label{sec:4}

We can promote functions in the 7D metric to be functions of $r$ to get
\begin{equation}
    ds^2=e^{2f(r)}(-dt^2+dr^2) + \lambda(r)^2ds^2_\text{Wu} \, .
\end{equation}
Define
\begin{equation}
    H(\lambda)=-\frac{m\lambda}{2}-\frac{3}{2m\lambda}+\frac{6}{m^3\lambda^3}\, .
\end{equation}
If we assume that the spinor $\epsilon$ satisfies
\beq
\td\gamma_r \,\epsilon = \epsilon \, ,
\label{proj}
\eeq
the gravitino variation then gives
\bea
    e^{-f(r)}\lambda'(r) & = & H(\lambda) \, , \\
    e^{-f(r)}f'(r) & = & H'(\lambda)\,,
\eea
where in all cases, a prime denotes a derivative with respect to the indicated argument.
Dividing these equations gives
\begin{equation}
\frac{df}{d\lambda}=\frac{H'}{H} \, ,
\end{equation}
which can be integrated with respect to $\lambda$ to give
\begin{equation}
    H(\lambda)=-e^{f(\lambda)} \, .
\end{equation}
Changing variables in the metric from $r$ to $\lambda$ then gives
\begin{equation}
\label{extbh}
    ds^2=-e^{2f}dt^2+e^{-2f}d\lambda^2+\lambda^2ds^2_\text{Wu} \, .
\end{equation}

The Einstein equations in the Wu directions are
\begin{equation}
    R_{ab}=\left(-e^{-2f}\lambda^{-1}f'\lambda'-e^{-f}\lambda^{-1}(\lambda'e^{-f})'-4e^{-2f}\lambda^{-2}\lambda'^2+\frac{6}{\lambda^2}\right)\delta_{ab}\,,
\end{equation}
and in the AdS$_2$ directions are
\begin{equation}
R_{00}=-R_{11}=e^{-2f}f''+5e^{-2f}\lambda^{-1}f'\lambda' \, ,
\end{equation}
where a prime here denotes a derivative with respect to $r$.

Switching to H as a function of $\lambda$,
\begin{equation}
\begin{split}
    R_{ab}=&\left( -\frac{2H(\lambda)H'(\lambda)}{\lambda}-4\frac{H(\lambda)^2}{\lambda^2}+\frac{6}{\lambda^2} \right) \delta_{ab}\\
    =&\left(-\frac{3}{2}m^2+\frac{15}{2m^2\lambda^4}+\frac{72}{m^6\lambda^8} \right) \delta_{ab} \, ,
\end{split}
\end{equation}
and
\begin{equation}
\begin{split}
    R_{00}=-R_{11}=&H(\lambda)H''(\lambda)+H'(\lambda)^2+\frac{5H(\lambda)H'(\lambda)}{\lambda}\\
    =&\frac{3}{2}m^2+\frac{15}{2m^2\lambda^4}+\frac{216}{m^6\lambda^8} \, .
\end{split}
\end{equation}
Since these are consistent with the results of section \ref{sec:Eins}, the solution \eq{extbh} satisfies both the BPS and Einstein equations, and preserves 1/32 supersymmetry. Note that the supersymmetry is halved compared to the AdS$_2 \times W$ product solution because we imposed an additional constraint, eqn \eq{proj}. We will see in the next section that the solution \eq{extbh} describes the extremal case of the black hole solution \eq{eq:BH}. 

\subsection{Generalisation to a black hole solution}

  A somewhat analogous system to the one we are investigating in this paper 
was studied in \cite{cvlupoYM}, where a non-abelian black hole was constructed as a solution of five-dimensional gauged supergravity.  In that case, the theory was shown to admit a solution of the form AdS$_2\times S^3$, in which the $SU(2)$ gauge fields of the theory were taken to be given by the natural connection on $S^3$.  It was shown in \cite{cvlupoYM} that, as well as the AdS$_2\times S^3$ solution, which generalises in our present case to AdS$_2\times W$, the solution could be extended by adding in a black hole mass term.  

In our case, we can also obtain a more general solution, in which a mass term is added.  Thus, we find that
the following solution arises:
\bea
\label{eq:BH}
ds_7^2 &=& -h(r)\, dt^2 + h(r)^{-1}\, dr^2 + r^2\, d s_{\rm Wu}^2\,,\nn\\
G^{AB} &=& \fft1{2m}\, F^{AB}\,,\nn\\
S^A &=& -\fft{2\sqrt3}{m^4\, r^4}\, dt\wedge dr\wedge e^A\,,
\eea
with
\bea
h(r)= \fft32 + \fft{36}{m^6\, r^6} -\fft{15}{4 m^2\, r^2} + \fft14 m^2\, r^2 - \fft{\alpha}{r^4}\,,
\eea
where $\alpha$ is an arbitrary constant (like a mass parameter in a black hole solution).

Note that if $\alpha$ is taken to equal $\dfft{18}{m^4}$, the metric function $h$ becomes
\bea
h(r)\Big|_{\alpha=\fft{18}{m^4}} = \fft{(m^4\, r^4 + 3 m^2\, r^2 -12)^2}{4 m^6\, r^6}\,.\label{hsol}
\eea
The numerator here is proportional to the quadratic function we encountered in eqn (\ref{quadratic}) (with $\lambda$ in that equation now replaced by $r$).  Thus, we could choose to write the metric function as
\bea
h(r)= \fft{(m^4\, r^4 + 3 m^2\, r^2 -12)^2}{4 m^6\, r^6} -\fft{\td\alpha}{r^4}\,,\qquad \hbox{where}\qquad \td\alpha = \alpha - \fft{18}{m^4}\,.
\eea

It is interesting to note 
that there exists a solution matching the setup of \cite{Birmingham} also if the Yang-Mills fields are turned off.  This has the effect of
removing the second and third terms on the right-hand side of eqn (\ref{hsol}), giving
\bea
h(r)= \fft32 + \fft14 m^2\, r^2 - \fft{\alpha}{r^4}\,,
\eea
By making constant rescalings to the new quantities $\td t= \sqrt{\ft32}\, t$,\ $\td r= \sqrt{\ft23}\, r$ and
$d\td s_5^2 = \ft32\, ds^2_\text{Wu}$, the metric takes the form of a Schwarzschild-AdS$_7$ black hole in seven dimensions, except with the usual 5-sphere replaced by the rescaled Wu metric $d\td s_5^2$ with the same
Einstein constant as the unit 5-sphere, i.e. $\wtd R_{ab}= 4 \td g_{ab}$.

\subsection{Properties and thermodynamics of the black hole}

Consider the black hole
\begin{equation}
\begin{split}
    &ds_7^2=-h(r)dt^2+h^{-1}(r)dr^2+r^2ds^2_\text{Wu} \, , \label{blackhole}\\
    \text{where }&h(r)=\frac{(m^4r^4+3m^2r^2-12)^2}{4m^6r^6}-\frac{\tilde{\alpha}}{r^4} \, ,
    \end{split}
\end{equation}
with $r$ in the physical range $r>0$. 

For $\tilde{\alpha}<0$, there are no zeros of $h$ and therefore no horizons. 

For $\tilde{\alpha}=0$, there exists one root of $h$ for which $h(r_0)=0$ defined by $m^2r^2=m^2r_0^2=\frac{\sqrt{57}-3}{2}$ where also $h'(r_0)=0$. Therefore, $\tilde{\alpha}=0$ defines an extremal BPS black hole, which is the solution in eqn \eq{extbh}.  In the near-horizon limit, the extremal metric approaches the AdS$_2\times \,$(Wu manifold) solution found in section \ref{sec:3}.

Consider when $\tilde{\alpha}>0$. When $r$ tends to $0$ from above, $h$ tends to $+\infty$. As $r\rightarrow \infty$, $h\rightarrow \infty$. Therefore, the lowest root of $h(r)$ has a negative derivative $h'(r)<0$, followed by alternating signs of the derivative until the last root, which has a positive derivative.

At a point $r=\tilde r$ where $h(\tilde r)=0$, the derivative at that point is
\begin{equation}
    h'(\td r)=\frac{3(4+m^2\td r^2+m^4 \td r^4)(-12+3m^2\td r^2+m^4\td r^4)}{2m^6 \td r^7} \, .
\end{equation}
At a root $\td r<r_0$, the derivative is negative, so there is only one root of $h(r)$ for $r<r_0$. At a root $\td r>r_0$, the derivative is positive, so there is only one root of $h(r)$ for $r>r_0$.  The conclusion is that there are two positive real roots in the interval $0\le r\le \infty$ whenever $\tilde\alpha>0$, and no real roots when $\tilde \alpha<0$, can also be seen by applying Sturm's theorem. 

For $\tilde{\alpha}>0$, there is an inner horizon $r_-$ and an outer horizon $r_+$. The Hawking temperature is defined by
\begin{equation}
    T=\frac{h'(r_+)}{4\pi}=\frac{3(4+m^2r_+^2+m^4r_+^4)(-12+3m^2r_+^2+m^4r_+^4)}{8\pi m^6r_+^7} \, .
\end{equation}
The entropy is then given by
\begin{equation}
    S=\frac{\text{Vol(Wu)}r_+^5}{4G_7}\, , \label{ent}
\end{equation}
where $\text{Vol(Wu)}$ is the volume of the Wu manifold (calculated in our conventions in \cite{Wu})
\begin{equation}
\text{Vol(Wu)}=\frac{\sqrt{3}\pi^3}{8}\, . \label{vol}
\end{equation}

The heat capacity is then given by
\begin{equation}
    T\frac{\frac{\partial S}{\partial r_+}}{\frac{\partial T}{\partial r_+}}=\frac{5r_+^5(4+m^2r_+^2+m^4r_+^4)(-12+3m^2r_+^2+m^4r_+^4)\,\text{Vol(Wu)}}{4G_7\,(336+15m^4r_+^4-4m^6r_+^6+m^8r_+^8)} \, .
\end{equation}
This is positive at the outer horizon $r=r_+$, indicating local thermodynamic stability.

Since the black hole has zero angular momentum, if we keep the charge associated with the nonabelian gauge field fixed, we can integrate $dE=TdS$ to get the energy above the ground state energy $E_0$ corresponding to the extremal case:
\begin{equation}
    E-E_0=\frac{5\,\text{Vol(Wu)}\tilde{\alpha}}{16\pi G_7\, } \, .
\end{equation}

The quantum statistical relation here, simplified since there is no angular momentum, is
\begin{equation}
\begin{split}
    I=&-S+\beta (E-E_0)\\
    =&-\frac{\sqrt{3}\pi^3r_+^5}{192G_7}\left(\frac{m^4r_+^4-9m^2r_+^2+84)}{4+m^2r_+^2+m^4r_+^4}\right) \, .
\end{split}
\end{equation}

\subsection{Holography}

Via the AdS/CFT correspondence \cite{malda,gubklepol,withol}, AdS$_7$ black holes are expected to be dual to a $6D$ $\mathcal{N}=(2,0)$ SCFT (see, for example, \cite{bobev1,bobev2} for recent work on holography in this context and \cite{holreview} for a review). In our case, eqn (\ref{blackhole}) describes a black hole asymptotic not to AdS$_7$ but to AdS$_7$ with Minkowski slices replaced by the product of time and the Wu manifold. This can be viewed as being dual to a $6D$ $(2,0)$ SCFT in the UV. Furthermore, the near-horizon limit is AdS$_2\times W$. This can be viewed as being dual to a 1D SCFT in the IR. Something interesting in this case is that here the theory is twisted by an $SO(3)$ gauge field. Therefore, one should, in principle, be able to calculate a twisted topological index in the dual theory. Similar to \cite{Douglas:2010iu}, one can consider a 6D $(2,0)$ SCFT on $W \times S^1$, then compactify on the circle to consider MSYM on Wu twisted by $SO(3)$ gauge fields embedded in the $SO(5)$ R-symmetry.

Note that our solution can be interpreted as a wrapped fivebrane in the sense that the product of the Wu manifold and a real line (or circle in the case of Euclideanised time) can be viewed as the worldvolume of a fivebrane.

The entropy from eqn \eq{ent} can be written using the explicit volume in eqn \eq{vol} as
\begin{equation}
    S=\frac{\sqrt{3}\pi^3r_+^5}{32G_7} \, .
\end{equation}
We can use the standard holographic dictionary (see, for example, \cite{bobev1}) to write
\begin{equation}
    N^3=\frac{3\pi^2L_{\text{AdS}_7}^5}{16G_7} \, ,
\end{equation}
where the length scale of AdS$_7$ is $L_{\text{AdS}_7}=\frac{2}{m}=\frac{4}{g}$.
So the entropy can be written as
\begin{equation}
    S=\frac{ \sqrt{3}\pi}{192}m^5r_+^5N^3 \, ,
\end{equation}
giving the usual $N^3$ scaling for entropy in AdS$_7$/CFT$_6$ \cite{Klebanov:1996un,Henningson:1998gx}.

\subsection{An analytic flow to AdS$_2\times H^3 \times H^2$}

Another example of a flow between an AdS$_2$ times a 5-manifold IR and an AdS$_7$ type UV is the construction in \cite{Gauntlett:2001jj} using the 5-manifold $H^3\times H^2$, the product of hyperbolic spaces of dimension $2$ and $3$. In appendix \ref{apph2h3}, the full derivation is given. 

\section{Conclusions}
\label{sec:5}

In this paper, we compactified 7D gauged maximal supergravity on the
$SU(3)/SO(3)$ Wu manifold, and its noncompact dual $SL(3,\R)/SO(3)$ coset. The Wu manifold is a spin$^h$ manifold that is neither spin nor spin$^c$, yet we used the gauge group of the 7D theory to find supersymmetric (1/16 BPS) AdS$_2 \times W$ solutions. Since the 7D theory can be obtained via an $S^4$ reduction of $D=11$ supergravity, we can uplift the solution to an 11D AdS$_2 \times M_9$ solution, where $M_9$ is an $S^4$ bundle over the Wu manifold, with the fibres twisted by an $SO(3)$ gauge field. This 9-manifold $M_9$ is the unit tangent bundle of the Wu manifold, and we investigated its topological properties while showing how this construction fits into previous classifications of AdS$_2$ $D=11$ supergravity solutions \cite{zayas,bubbling}. Later, we generalised the AdS$_2 \times W$ solution to an extremal AdS$_7$-like non-Abelian black hole solution with 1/32 supersymmetry and a nonsupersymmetric one.

In \cite{Gauntlett:2000ng}, solutions AdS$_2\times M_5$ for 5-manifolds $S^5$ and $H^5$ were found in gauged seven-dimensional supergravity. Here, we have obtained solutions for AdS$_2$ times $SU(3)/SO(3)$ and $SL(3)/SO(3)$. Together, these four five-dimensional spatial factors comprise all the irreducible symmetric spaces. A natural question to ask if there is a general construction of this type of supergravity solution using symmetric spaces.

It is interesting to compare the AdS$_2\times W$ BPS solution with the lack of an AdS$_3\times\mathbb{CP}^2$ BPS solution \cite{Gauntlett:2000ng}, especially since that would seem to be a natural spin$^c$ analogy of our spin$^h$ construction here. In \cite{Wu}, we explicitly calculated an $SO(3)$-gauge covariantly constant spinor\footnote{Technically, this spinor is charged under the $U(2)=(SU(2)\times U(1))/\Z_2$ as a singlet under $U(1)$ and with $SO(3)$ isospin $j=\frac{1}{2}$ (for spin$^h$ spinors, $SO(3)$ has half-integer isospin as in footnote \ref{footnote}).} on $\mathbb{CP}^2$. It would be interesting to see if this charged spinor could be used in similar gauged supergravity solutions as the Wu manifold, as well as seeing if more generally gauge-covariantly constant spinors could be more systematically utilised in supergravity solutions. 

It would be interesting, although very involved, to do a careful computation of the twisted topological index of the dual 6D SCFT, perhaps by doing a circle compactification of the theory defined on $W \times S^1$ to consider MSYM on the Wu manifold.

In \cite{Gauntlett:2001jj}, BPS equations were formulated for a flow to an AdS$_2\times H^3\times H^2$ fixed point from an AdS$_7$-type fixed point. In this paper, we gave an analytic solution for the BPS flow between these fixed points using changes of variables suggested by the fixed point solutions. It would be interesting to investigate whether there are similar unsolved BPS equations that are analytically tractable using similar methods.

\section*{Acknowledgments}
The authors would like to thank Fri{\dh}rik Freyr Gautason for helpful discussions about the holographic interpretation of the solution presented in this paper and Arun Debray for helpful discussions about the topology of the twisted $S^4$ bundle over the Wu manifold. The authors would also like to thank the University of Cambridge for hospitality while part of this work was being carried out.

\bigskip

\appendix

\section{The Wu metric, and its $SO(3)$ Yang-Mills field strengths}
\label{wurecap}

Here, we briefly summarise the metric of the Wu manifold and give explicit expressions for the principal $SO(3)$ bundle, 
as in \cite{Wu}.
The Wu metric can be written as
\bea
\label{WuMetric}
ds_5^2 = d\mu^2+ \ft13 d\nu^2 + \sin^2(\mu-\nu)\,\sigma_1^2 +
  \sin^2(\mu+\nu)\, \sigma_2^2 + \sin^2 2\mu\, \sigma_3^2\,,
\eea
and its $SO(3)$ gauge fields are
\bea
A^1=\cos(\mu-\nu)\, \sigma_1\,,\qquad
A^2= \cos(\mu+\nu)\, \sigma_2\,,\qquad
A^3=\cos 2\mu\, \sigma_3\,,\label{so3YM}
\eea
where the left-invariant 1-forms $\sigma_i$ are defined on the 
lens space $S^3/D_8^* = SO(3)/D_4$ and $\mu\in [0,\pi/2]$, $\nu\in[-\mu,\mu]$.
We shall use the natural vielbein basis $e^a$, with
\bea
e^1= \sin(\mu-\nu)\, \sigma_1\,,\quad
e^2= \sin(\mu+\nu)\, \sigma_2\,,\quad 
e^3=\sin2\mu\, \sigma_3\,,\quad
e^4=d\mu\,,\quad e^5=\fft{d\nu}{\sqrt3}\,.\label{wuviel}
\eea

The $SO(3)$ Yang-Mills field strengths 
$F^i=dA^i+\ft12\ep_{ijk}\, A^j\wedge A^k$ are given by
\bea
F^1 &=& e^1\wedge e^4-e^2\wedge e^3 - \sqrt3\, e^1\wedge e^5\,,\nn\\
F^2 &=& e^2\wedge e^4 + e^1\wedge e^3 +\sqrt3\, e^2\wedge e^5\,,\nn\\
F^3 &=& 2 e^3\wedge e^4 + e^1\wedge e^2\,.\label{SO3Fi}
\eea
The field strengths $F^i$ give the curvature of the principal $SO(3)$ bundle.
Note that they obey
\bea
F^i_{ac}\, F^i_{bc}= 6\delta_{ab}\,,\qquad
F^i_{ab}\, F^j_{ab}=10\delta_{ij}\,.\label{FFid}
\eea
They also satisfy the Yang-Mills equations $D{\ast F}^i=0$, and in fact they have the stronger property of being gauge-covariantly constant:
\bea
D_a\, F^i_{bc} \equiv \nabla_a\, F^i_{bc} + \epsilon_{ijk}\, A^j_a\, F^k_{bc}=0\,.
\eea

\section{Kaluza-Klein description of the 9-manifold}
\label{kkcalc}
In the 9D vielbein basis, the Wu manifold has $R_{ab}=\frac{6}{\lambda^2}\delta_{ab}$. $S^4$ has radius $\frac{1}{m}$ so $R_{\alpha \beta}=3m^2\delta_{\alpha \beta}$. Switching to the 9D vielbein basis gives $\mathcal{F}^i_{ab}=\frac{1}{\lambda^2}F^i_{ab}$. Also, since $D_bF_a^{ib}=0$, $R_{a\alpha}=0$. Using these, we get

\begin{equation}
    \begin{split}
        &\hat{R}_{ab}=\frac{6}{\lambda^2}\delta_{ab}-\frac{1}{2}K^{\alpha i}K_\alpha^j\mathcal{F}_{a c}^i\mathcal{F}_b ^{jc} \, ,\\
        &\hat{R}_{a \alpha}=0 \, ,\\
        &\hat{R}_{\alpha \beta}=3m^2\delta_{\alpha \beta}+\frac{5}{2\lambda^4}K_\alpha^iK_\beta^i \, .
    \end{split}
\end{equation}
We can therefore calculate the Ricci scalar (using $F^i_{ab}F^{jab}=10\delta^{ij}$ again and $F^i_{ab}F^i_{bc}=6\delta_{ac}$) to get
\begin{equation}
    \begin{split}
        \hat{R}&=\frac{30}{\lambda^2}-\frac{1}{2\lambda^4}K^{\alpha i}K^j_\alpha 10\delta^{ij}+12m^2+\frac{5}{2\lambda^4}K^i_\alpha K^i_\alpha\\
        &=12m^2+\frac{30}{\lambda^2}-\frac{5}{2\lambda^4}\frac{1}{m^2}(Y^T{T^{i}}^TT^iY)\\
        &=m^2 (12+ 30 C^2 - 15 C^4) \, .
    \end{split}
\end{equation}

Although $\hat{R}_{ab}\hat{R}^{ab}$ is constant, $\hat{R}_{ab}\hat{R}_{bc}\hat{R}_{ac}$ depends on coordinates, showing that the 9-manifold is non-homogeneous.

\section{$SU(4)$ conditions}
\label{pz}

The conditions for 11D supersymmetry from \cite{zayas} that we satisfy are
\begin{subequations}
\begin{align}
&d(e^A J)=0 \, ,\label{eq:diffcond1}\\
&d(e^{2A}V) + e^{A}J+ e^{2A} G_2=0 \, ,\label{eq:diffcond2}\\
&d(e^A V\wedge \text{Im}\Omega)-e^{A} J\wedge G_4=0 \, ,\label{eq:diffcond3}\\
&\ast_9\big(2V\wedge \ast_9 G_2+  \text{Re}\Omega\wedge G_4\big)+6 dA=0 \, ,\label{eq:diffcond4}\\
&d(e^{2A} \text{Re}\Omega)- e^{A} V\wedge \text{Im}\Omega+ e^{2A}(\ast_9 G_4-V\wedge G_4)=0 \, ,\label{eq:diffcond5}
\end{align}
\end{subequations}
together with the algebraic constraints
\begin{subequations}
\begin{align}
&J\wedge J\wedge G_4=0 \, ,\label{eq:algebcond1}\\
&V\wedge (\text{Im}\Omega \wedge G_2+ J\wedge G_4)=0 \, ,\label{eq:algebcond2}\\
&e^{A}\big(2J\wedge \ast_9 G_2-  V\wedge \text{Im}\Omega\wedge G_4\big)=-6 \text{Vol}_9\label{eq:algebcond3} \, ,
\end{align}
\end{subequations}
Note that the right-hand side of the last constraint differs by a sign from \cite{zayas}, perhaps a typo.

\section{The eleven-dimensional lift of the AdS$_2\times S^5$ solution\label{S4overS^5}}

The AdS$_2\times S^5$ solution of seven-dimensional gauged supergravity was obtained in \cite{Gauntlett:2000ng}.  This can be lifted to an AdS$_2\times M_9$ solution of eleven-dimensional supergravity, for which $M_9$ is of the form of the unit tangent bundle of $S^5$, that is, $M_9=ST(S^5)$. This 9-manifold is in fact the homogeneous Stiefel manifold
$SO(6)/SO(4)$. If we introduce left-invariant 1-forms $M_{AB}$ where $1\le A\le 6$ and $M_{AB}=-M_{BA}$, obeying the Maurer-Cartan algebra $d M_{AB}= M_{AC}\wedge M_{CB}$, then we can write the family of metrics $ds_9^2$ on
$M_9$, with\footnote{Note that because of a shortage of convenient alphabets, some of our choices of indices in our
discussion here are specific just to this appendix.}
\bea
ds_9^2= M_{a i}^2 + a^2\, M_{56}^2\,.\label{Stiefelmetric}
\eea
Here, we have decomposed the $SO(6)$ indices as $A=(a,i)$, with $1\le a\le 4$ and $5\le i\le5$.  
The constant $a$ in eqn (\ref{Stiefelmetric}) is a squashing parameter.  In this example, unlike the case of the unit tangent bundle of the Wu manifold that we were studying in the bulk of the paper, $M_9$ is
actually a $U(1)$ bundle over a smooth 8-dimensional manifold $M_8$, whose metric is 
\bea
ds_8^2 = M_{ai}^2\,.
\eea
Note, in particular, that the vector dual to $M_{56}$ is a Killing vector in the nine-dimensional metric $ds_9^2$, and the relation
$ds_9^2 = ds_8^2 + a^2\, M_{56}^2$ is of the form of a standard Kaluza-Klein $U(1)$ reduction.

 The reduced 8-dimensional manifold $M_8$ is itself the Grassmannian coset space
\begin{equation}
    G_2^+(\R^6)=\frac{SO(6)}{SO(2)\times SO(4)},
\end{equation}
which is isometric to 
\begin{equation}
    G_2(\C^4)=\frac{U(4)}{U(2)\times U(2)}.
\end{equation}
This manifold is discussed in the context of supergravity solutions in appendix A.4. of \cite{fof}.

   Let us introduce the vielbein $e^a$ for $ds_9^2$, with
\bea
e^1&=& M_{15}\,,\quad e^2=M_{16}\,,\quad e^3= M_{25}\,,\quad e^4=M_{26}\,,\nn\\
e^5 &=& M_{35}\,,\quad e^6=M_{36}\,,\quad e^7 = M_{45}\,,\quad e^8=M_{46}\,,\nn\\
\quad e^9 &=& a\, M_{56}\,.
\eea
It can be verified that the Ricci tensor of the 9-dimensional metric $ds_9^2$ has vielbein components
\bea
^{(9)}R_{\alpha\beta} = (4-\ft12 a^2)\, \delta_{\alpha\beta}\,,
\qquad ^{(9)}R_{99}= 2 a^2\,,\quad \hbox{where}\quad 1\le\alpha\le 8
\,,
\eea
and with all other components vanishing, and that the 8-dimensional metric $ds_8^2$ has the Ricci tensor
\bea
^{(8)}R_{\alpha\beta}=4\delta_{\alpha\beta}\,.
\eea

Defining $V=e^9$, so that $ds_9^2 = ds_8^2 + V^2$, we have that $J= -a^{-1}\, dV$, with
\bea
J= e^1\wedge e^2 + e^3\wedge e^4 + e^5\wedge e^6 + e^7\wedge e^8\,.\label{Js5}
\eea
It is associated with an integrable complex structure (one can verify straightforwardly that the Nijenhuis tensor vanishes), and in 
fact $J$ is the K\"ahler form on $M_8$.  It obeys $dJ=0$ and $d{\,\ast_8\, J}=0$.  

We also introduce the complex 1-forms $Z^a$ and their complex conjugates $\bar Z^a$: 
\bea
Z^a \equiv  M_{a5} + \im M_{a6}\,,\qquad \bar Z^a= M_{a5} -\im M_{a6}\,.
\eea
We can define the 4-form $\omega_4$:
\bea
\omega_4 = \fft{1}{8} \epsilon_{abcd} Z^a\wedge Z^b\wedge \bar Z^c\wedge \bar Z^d\,.
\eea
This 4-form is real, and, with respect to the complex structure defined by $J$, it is of type $(2,2)$.
We have
\bea
\omega_4 &=& 3 e^{1357} + e^{1467} + e^{2367} + e^{2457} + 3 e^{2468} + e^{1368} + e^{1458} + e^{2358}\,,
\eea
where $e^{ijkl}$ means $e^i\wedge e^j\wedge e^k\wedge e^\ell$.
It can be verified that 
\bea
d\omega_4=0\,,\qquad \omega_4=\ast_8\, \omega_4\,,\qquad \omega_4\wedge \omega_4 = 24 \text{Vol}_8\,.\label{om4s5}
\eea
Thus $\omega_4$ is harmonic in $M_8$, and in fact it is 
one harmonic representative of the cohomology class $H^4(M_8;\R)=\R \oplus \R$
on $M_8$ (see, for example, \cite{lu2023nonlinear}). It is primitive, meaning that 
\bea
J\wedge\omega_4=0\,.\label{om4prim}
\eea
The second independent harmonic representative is $J\wedge J$.

Additionally, we may define the holomorphic 4-form $\Omega_4$ as
\bea
\Omega_4= Z^1\wedge Z^2\wedge Z^3\wedge Z^4\,.
\eea
This is of type $(4,0)$.
It may be verified that the following identities hold:
\bea
d\Omega_4 = -\fft{4\im}{a}\, V\wedge \Omega_4\,,\qquad \omega_4\wedge \Omega_4=0\,.
\eea
We also have
\bea
\ast_8\, J &=& \fft16 J\wedge J\wedge J\,,\qquad \ast_9\, J= \fft16 V\wedge J\wedge J\wedge J\,,\nn\\
\ast_9\, \omega_4 &=& V\wedge (\ast_8\, \omega_4)= V\wedge \omega_4\,.
\eea
Note that $d\,{\ast_9 \,\omega_4}= dV\wedge \omega_4= -\fft1{a}\, J\wedge\omega_4=0$ since $\omega_4$ is primitive 
(see eqn(\ref{om4prim})), so $\omega_4$ is also harmonic in $M_9$; it is the harmonic representative of the4th cohomology class $H^4(M_9;\R)=\R$.  Note that $J\wedge J$ does not provide a second independent representative of $H^4(M_9;\R)$, since in $M_9$ $J\wedge J$ is exact; it can be written as $d(-a^{-1}\, V\wedge J)$, and $V\wedge J$
is globally defined in $M_9$ (but not in $M_8$).  This is consistent with the fact that in $M_8$ we have $H^4(M_8;\R)= \R\oplus \R$, but in
$M_9$ we just have $H^4(M_9;\R)=\R$.

Using all of the above properties, together with eqns (\ref{G2G4eom}), one can verify that by taking the eleven-dimensional metric and 4-form field strength
to be given by
\begin{equation}
\begin{split}
&ds_{11}^2=\lambda^2(e^{2A}d\bar{s}^2_{\text{AdS}_2}+ds_9^2) \, ,\\
&F_4= \lambda^{3}\, (G_4+e^{2A}\overbar{\text{Vol}}_{\text{AdS}_2}\wedge G_2) \,,
\end{split}
\end{equation}
and with $a=1$ in eqn (\ref{Stiefelmetric}), 
\bea
G_2=-3 J\,,\qquad   G_4= \omega_4\,,\qquad  e^{-A}= 4\,,\label{G2G4S5}
\eea
a solution of eleven-dimensional supergravity is obtained. This is precisely the lift of the seven-dimensional AdS$_2\times S^5$
supergravity solution found in \cite{Gauntlett:2000ng}, and it is in accordance with the equations obtained in \cite{zayas}, which we reproduce in appendix \ref{pz}.  It can be verified that the conditions set out in section 2.4 of \cite{zayas} for having an
${\cal N}=(2,0)$ K\"ahler structure are satisfied, and we shall indeed confirm below that the supersymmetry preserved in this background is consistent with this.  It is interesting to note that the structure of the 4-form $G_4=\omega_4$, given in our eqn (\ref{om4s5}), is
not of the form described in eqn (2.45) in \cite{zayas} (even though we have the same canonical basis for $J$ as is used in \cite{zayas} (see our eqn (\ref{Js5})).

Note that with the constant $a$ in eqn (\ref{Stiefelmetric}) being set
to $a=1$, as is required by the $D=11$ supergravity equations of motion, the terms in the metric on the Stiefel 
manifold can be grouped as
\bea
ds_9^2 = \sum_{a=1}^5 M_{a6}^2 + \sum_{a=1}^4 M_{a5}^2\,,\label{Stiefelmetric2}
\eea
with the first term here being the metric on the unit-radius $S^5$ base, and the second term being the
metric on the unit-radius $S^4$ fibres.

It is straightforward to verify directly the supersymmetry of this Stiefel manifold solution of eleven-dimensional
supergravity.  In the nine internal directions, it follows from eqn (\ref{D11susy}) that we shall have
\bea
\delta\hat\psi_a &=& \hat\nabla_a\hat\ep -
   \ft18\oneone\otimes (3 \boldsymbol{J}\,\wtd\Gamma_a -\wtd\Gamma_a\,\boldsymbol{J})\,\hat\ep +
    \ft1{24} \td\gamma_\ast \otimes(3 \boldsymbol{\omega}_4 - \wtd\Gamma_a\, \boldsymbol{\omega}_4)\,\hat\epsilon
      \,,\qquad a=1\cdots 8\,,\nn\\
\delta \hat \psi_9 &=& \hat \nabla_9 \,\hat\epsilon - 
                 \ft18 \oneone\otimes (3\boldsymbol{J}\,\wtd\Gamma_9 -\wtd\Gamma_9\, \boldsymbol{J})\,\hat\epsilon
                  + \ft1{24} \td\gamma_\ast\otimes (3\boldsymbol{\omega}_4\,\wtd\Gamma_9 - \wtd\Gamma_9\,
                                   \boldsymbol{\omega}_4)\,\hat\epsilon\,,\label{D11susycon}
\eea
where we have defined the boldface quantities
\bea
\boldsymbol{J}= \ft12 J_{ab}\, \wtd\Gamma^{ab}\,,\qquad \boldsymbol{\omega}_4 =\ft1{24} \omega_{abcd}\, 
   \wtd\Gamma^{abcd}\,.
\eea
We can describe the metric (\ref{Stiefelmetric2}) by writing a family of homogeneous squashed metrics on the 
15-dimensional group manifold $SO(6)$ as
\bea
ds_{15}^2 = ds_9^2 + b^2\, \sum_{1\le a< b\le 4} M_{ab}^2\,.
\eea
The metric $ds_9^2$ is obtained by taking the Gromov-Hausdorff limit $b\longrightarrow0$. 

The spin connection must be handled a little carefully in this construction.  A straightforward calculation shows that 
after taking the Gromov-Hausdorff
limit, the spinor exterior derivative $\nabla=d + \ft14 \omega_{ab}\wtd \Gamma^{ab}$ in the 9-manifold is given by
\bea
\nabla &=& d - \ft14 J_{ab}\, \wtd\Gamma_{b9}\, e^a -\ft14 \boldsymbol{J}\, e^9 \nn\\
&&-(\wtd\Gamma_{13}+\wtd\Gamma_{24})\, M_{12} -  (\wtd\Gamma_{15}+\wtd\Gamma_{26})\, M_{13} 
 -(\wtd\Gamma_{17}+\wtd\Gamma_{28})\, M_{14}\nn\\
&&- (\wtd\Gamma_{35}+\wtd\Gamma_{46})\, M_{23} 
  -(\wtd\Gamma_{37}+\wtd\Gamma_{48})\, M_{24} - (\wtd\Gamma_{57}+\wtd\Gamma_{68})\, M_{34}\,.\label{spincon1}
\eea
The terms in the last two lines are proportional to the left-invariant 1-forms of the $SO(4)$ denominator group
in $M_9=\dfft{SO(6)}{SO(4)}$, and in fact, as we shall see below, they drop out when the supersymmetry conditions in eqns (\ref{D11susycon}) are
imposed. That is to say, from the terms in (\ref{spincon1}) on the first line, which lie in the nine directions in $M_9$,
we find that $\hat\nabla_a$ and $\hat\nabla_9$ in eqns (\ref{D11susycon}) are given by
\bea
\hat\nabla_a= \del_a -\ft14 J_{ab}\, \wtd\Gamma^b\,\wtd\Gamma_9\,,\qquad 
\hat\nabla_9 = \del_9 - \ft14 \boldsymbol{J}\,.
\eea
using these expressions in eqns (\ref{D11susycon}), we find that for supersymmetry parameters of the form 
$\hat\epsilon=\zeta\otimes\eta$, the 16-component spinor $\eta$ in $M_9$, taken to be independent of the
coordinates, must satisfy the projection conditions
\bea
J_{ab}\,\wtd\Gamma^b\,\eta - \ft14 \wtd\Gamma_a\,\boldsymbol{J}\,\eta =0\,.
\eea
These conditions imply that $\eta$ has just two independent components, taking the form 
\bea
\eta = (0,0,\eta_3,0,0,-\eta_3,0,0,\eta_9,0,0,0,0,0,0,-\eta_9)^T\,.
\eea
Such spinors then automatically satisfy $\boldsymbol{\omega}_4\,\eta=0$ and $\boldsymbol{\omega}_4\, \wtd\Gamma_a\,\eta=0$.
Furthermore, they automatically obey the projection conditions $(\wtd\Gamma_{13}+\wtd\Gamma_{24})\,\eta=0$, etc., thus 
ensuring that the terms in the covariant derivative in eqn (\ref{spincon1}) lying in the $SO(4)$ denominator group 
directions are projected out.  
We also checked the integrability condition for setting eqn \eq{D11susycon} to zero and found that the spinor $\eta$ with constant components is the only one.
It is convenient to organise the two independent spinors in terms of $\eta_+$ and $\eta_-$,
as defined previously in eqns (\ref{etapm}).  These satisfy
\bea
\boldsymbol{J}\, \eta_+ = 4 \eta_-\,,\qquad \boldsymbol{J}\, \eta_- = -4 \eta_+\,.
\eea

In the AdS$_2$ directions there are two Killing spinors $\xi_+$ and two Killing spinors $\xi_-$, obeying
\bea
\nabla_\alpha\, \xi_\pm = \pm c\, \td\gamma_\alpha\, \xi_\pm\,.
\eea
These are not eigenstates of the two-dimensional chirality operator $\td\gamma_\ast$, but the linear combinations
$\zeta_\pm$ defined by
\bea
\zeta_\pm= \ft12 (\xi_+ \pm \xi_- )
\eea
are.  They obey
\bea
\nabla_\alpha\, \zeta_\pm = c\, \td\gamma_\alpha\, \zeta_\mp\,,\qquad  \td\gamma_\ast\, \zeta_\pm = \pm\, \zeta_\pm\,.
\eea
Thus, the supersymmetry conditions $\delta\hat\psi_a=0$ in the $a=1,\ldots,9$ internal directions are satisfied by
any spinor of the form
\bea
\hat\epsilon= \alpha_1\, \zeta_+\otimes \eta_+ + \alpha_2\, \zeta_+\otimes\eta_- + \alpha_3\, \zeta_-\otimes\eta_+
+ \alpha_4\,\zeta_-\otimes\eta_-\,,\label{ephatint}
\eea
for arbitrary constants $\alpha_i$.  (Note that there is also an implicit sum over the two independent $\zeta_\pm$ on AdS$_2$).

From eqns (\ref{D11susy}) and (\ref{FhatGG}), the 
supersymmetry condition $\delta\hat\psi_\alpha=0$ in the AdS$_2$ directions is given in eqn (\ref{extsusy}).
Plugging in the expressions for $G_2$ and $G_4$ for this case, as given in eqns (\ref{G2G4S5}), and the expression 
for $\hat\epsilon$ in (\ref{ephatint}), we find that
the condition $\delta\hat\psi_\alpha=0$ implies $\alpha_4=\alpha_1$ and $\alpha_3=\alpha_2$.  Thus we have supersymmetry when
\bea
\hat\epsilon = \alpha_1\, (\zeta_+\otimes \eta_+ + \eta_-\otimes \eta_-) + 
                  \alpha_2\,(\zeta_+\otimes \eta_- -\zeta_-\otimes\eta_+)\,,
\eea
with $\alpha_1$ and $\alpha_2$ being arbitrary constants.  Bearing in mind that there is also an implicit summation over the two
independent $\zeta_+$ and $\zeta_-$ spinors in AdS$_2$, this means the background preserves a fraction $\fft18$ of the original
32 components of the eleven-dimensional supersymmetry parameter.

\section{AdS$_2\times H^3 \times H^2$ flow}
\label{apph2h3}
Here, we give an explicit solution for the full BPS equations for a wrapped brane system described in \cite{Gauntlett:2001jj}, for which only the UV and IR fixed points were presented. Note that we follow the notation in \cite{Gauntlett:2001jj} and keep it self-contained to this section. In that paper, they consider an ansatz involving the direct product of $H^3$ with metric $d\bar{s}_1^2$ and $H^2$ with metric $d\bar{s}_2^2$
\begin{equation}
    ds_7^2=e^{2f}(-dt^2+dr^2)+e^{2g_1}d\bar{s}_1^2+e^{2g_2}d\bar{s}_2^2 \, .
\end{equation}
In this case, there is a scalar field turned on using
\begin{equation}\Pi_A^i=(e^{2\lambda},e^{2\lambda},e^{2\lambda},e^{-3\lambda},e^{-3\lambda}) \, ,
\end{equation}
and the 3-form potential is written as
\begin{equation}
    S_m=-\frac{l_1 l_2e^{-2g_1-2g_2+4\lambda+2f}}{4\sqrt{3}m^4}dt\wedge dr \wedge e^m \, ,
\end{equation}
for $m=1,2,3$.

The BPS equations to solve (eqn 4.5. in \cite{Gauntlett:2001jj}) are
\begin{equation}
\begin{split}
e^{-f}f'&= - {m\over 10}\left[3e^{-4\lambda}+2e^{6\lambda}\right]
+{3l_1 \over 20m }{e^{4 \lambda-2g_1}}
+{l_2 \over 10m }{e^{-6 \lambda-2g_2}}
-{9l_1l_2\over 20 m^3}{e^{2\lambda-2g_1-2g_2}}\\
e^{-f}{g_1'} &= - {m\over 10}\left[3e^{-4\lambda}+2e^{6 \lambda}\right]
- {7l_1 \over 20m }{e^{4 \lambda-2g_1}}
+ {l_2 \over 10m }{e^{-6 \lambda-2g_2}}
+{l_1l_2\over 20 m^3}{e^{2\lambda-2g_1-2g_2}}\\
e^{-f}{g_2'} &= - {m\over 10}\left[3e^{-4\lambda}+2e^{6 \lambda}\right]
+ {3l_1 \over 20m }{e^{4 \lambda-2g_1}}
- {2l_2 \over 5m }{e^{-6 \lambda-2g_2}}
+{3l_1l_2\over 10 m^3}{e^{2\lambda-2g_1-2g_2}}\\
e^{-f}\lambda' &=  {m\over 5}\left[ 
e^{6\lambda} - e^{-4\lambda}\right] 
+{l_1\over 10m } {e^{4\lambda-2g_1}}
-{l_2\over 10m } {e^{-6\lambda-2g_2}}
-{l_1l_2\over 20 m^3}{e^{2\lambda-2g_1-2g_2}} \, .
\end{split} \label{h2h3}
\end{equation}

Note that the fourth equation is somewhat redundant as dividing the first three equations by it gives three ODEs as functions of $\lambda$ without $f$ appearing. $f$ can then be calculated once the other functions are calculated.

There is a fixed point solution AdS$_2\times H^3 \times H^2$ presented in \cite{Gauntlett:2001jj} for $g_1=g_2$ and $l_1=l_2=-1$:
\begin{equation}
e^{10\lambda}=2 \, ,\quad 
e^{2g_i}={e^{8\lambda}\over 2 m^2} \, ,\quad
e^f={e^{4\lambda}\over 2m}{1\over r} \, .
\label{fixsol}
\end{equation}

Motivated by this solution, make the change of variables
\begin{equation}
    \Lambda=e^{10\lambda} \, , \quad G_i=\frac{l_i}{2m^2}e^{8\lambda-2g_i} \, , \quad \frac{d}{ds}=\frac{1}{m}e^{4\lambda-f}\frac{d}{dr}\, .
\end{equation}
This makes the fixed point solution \eq{fixsol} simply
\begin{equation}
    (\Lambda,G_1,G_2)=(2,-1,-1) \, .
\end{equation}

Under our changes of variables, the BPS equations become
\begin{equation}
    \begin{split}
        &\frac{d\Lambda}{ds}=2\Lambda\left(\Lambda-1+G_1- \frac{G_2(1+G_1)}{\Lambda}\right) \, ,\\
        &\frac{dG_1}{ds}=G_1\left(2\Lambda-1+3 G_1-\frac{2G_2(1+G_1)}{\Lambda}\right) \, ,\\
        &\frac{dG_2}{ds}=G_2\left(2\Lambda-1+G_1-\frac{4G_1G_2}{\Lambda}\right) \, .
    \end{split}
\end{equation}
Since $G_2$ always comes with a factor of $\frac{1}{\Lambda}$, define
\begin{equation}
    \Gamma_2=\frac{G_2}{\Lambda} \,,
\end{equation}
which transforms the BPS equations into
\begin{equation}
    \begin{split}
        &\frac{d\Lambda}{ds}=2\Lambda\left(\Lambda-1+G_1- \Gamma_2(1+G_1)\right) \, ,\\
        &\frac{dG_1}{ds}=G_1\left(2\Lambda-1+3 G_1-2\Gamma_2(1+G_1)\right) \, ,\\
        &\frac{d\Gamma_2}{ds}=\Gamma_2(1-G_1)(1+2\Gamma_2) \, .
    \end{split}
\end{equation}

There are four fixed points of these equations (keeping in mind that we should have $\Lambda=e^{10\lambda}>0$). $(\Lambda,G_1,\Gamma_2)=\left(2,-1,-\frac{1}{2}\right)$ is AdS$_2\times H^3\times H^2$. $(\Lambda,G_1,\Gamma_2)=\left(2,-1,0\right)$ is AdS$_4\times H^3$ since taking the limit $e^{-g_2}\rightarrow 0$ gives eqn~(3.5) of \cite{Gauntlett:2000ng}. $(\Lambda,G_1,\Gamma_2)=\left(\frac{1}{2},0,-\frac{1}{2}\right)$ is AdS$_5\times H^2$, which can be seen in that the BPS equations eqn~(\ref{h2h3}) are exactly equivalent to eqn~(23) of \cite{maldacenanunez} under the redefinitions $\lambda\rightarrow - \frac{1}{2}\lambda, \, l_2\rightarrow -\kappa, \, m\rightarrow 2$ and the limit $e^{-2g_1}\rightarrow 0$. $(\Lambda,G_1,\Gamma_2)=(1,0,0)$ is AdS$_7$-type in the same way as the solution with the Wu manifold was.

The flow we want to solve should link the two ends
\begin{equation}
(\Lambda,G_1,\Gamma_2)_\text{IR}=\left(2,-1,-\frac{1}{2}\right) \, , \quad (\Lambda,G_1,\Gamma_2)_\text{UV}=(1,0,0) \, .
\end{equation}
There is a solution linking the IR to the UV fixed points which takes a very simple form in these variables
\begin{equation}
    G_1=1-\Lambda, \quad \quad 2\Gamma_2=1-\Lambda \, .
\end{equation}
Then
\begin{equation}
    \frac{d\Lambda}{ds}=-\Lambda(\Lambda-1)(\Lambda-2) \, ,
\end{equation}
which integrates to
\begin{equation}
    e^{s-s_0}=\frac{\Lambda-1}{\sqrt{\Lambda(2-\Lambda)}} \, ,
\end{equation}
which rearranges to
\begin{equation}
    \Lambda(s)=1+\frac{1}{\sqrt{1+e^{-2(s-s_0)}}} \, .
\end{equation}
The whole solution can then be rewritten using $s$ as a coordinate:
\begin{equation}
    e^{2g_1}=\frac{l_1}{2m^2}\frac{\Lambda^{\frac{4}{5}}}{1-\Lambda} \, , \quad e^{2g_2}=\frac{l_2}{m^2}\frac{\Lambda^{-\frac{1}{5}}}{1-\Lambda} \, , \quad e^{2f}=e^{2f_0}\Lambda^{\frac{4}{5}}\frac{(2-\Lambda)^2}{\Lambda-1} \, .
\end{equation}
Then the 7-metric in these variables is
\begin{equation}
    ds_7^2=\Lambda^{\frac{4}{5}}\left(-e^{2f_0}\frac{(2-\Lambda)^2}{\Lambda-1}dt^2+\frac{1}{m^2}ds^2+\frac{l_1}{2m^2}\frac{1}{1-\Lambda}d\bar{s}_1^2+\frac{l_2}{m^2}\frac{1}{\Lambda(1-\Lambda)}d\bar{s}_2^2\right) \, ,
\end{equation}
with
\begin{equation}
    \Lambda(s)=1+\frac{1}{\sqrt{1+e^{-2(s-s_0)}}} \, .
\end{equation}


\bibliographystyle{JHEP}
\bibliography{biblio}

@phdthesis{Chen:2017,
  title={Bundles of irreducible Clifford modules and the existence of spin structures},
  author={Chen, Xuan},
  year={2017},
  school={State University of New York at Stony Brook},
  url={http://hdl.handle.net/11401/76378}
}

@article{HirzebruchHopf1958,
  author    = {Hirzebruch, Friedrich and Hopf, Heinz},
  title     = {Felder von {Fl\"achenelementen} in 4-dimensionalen {Mannigfaltigkeiten}},
  journal   = {Mathematische Annalen},
  volume    = {136},
  number    = {2},
  pages     = {156--172},
  year      = {1958},
  doi       = {10.1007/BF01362296},
  url       = {https://doi.org/10.1007/BF01362296}
}

@article{cvlupoYM,
    author = "Cvetic, Mirjam and Lu, H. and Pope, C. N.",
    title = "{Non-abelian black holes in D=5 maximal gauged supergravity}",
    eprint = "0908.0131",
    archivePrefix = "arXiv",
    primaryClass = "hep-th",
    reportNumber = "MIFP-09-35, UPR-1206-T",
    doi = "10.1103/PhysRevD.81.044023",
    journal = "Phys. Rev. D",
    volume = "81",
    pages = "044023",
    year = "2010"
    }

@article{Gauntlett:2000ng,
    author = "Gauntlett, Jerome P. and Kim, Nakwoo and Waldram, Daniel",
    title = "{M five-branes wrapped on supersymmetric cycles}",
    eprint = "hep-th/0012195",
    archivePrefix = "arXiv",
    reportNumber = "QMW-PH-00-16",
    doi = "10.1103/PhysRevD.63.126001",
    journal = "Phys. Rev. D",
    volume = "63",
    pages = "126001",
    year = "2001"
}

@article{Gauntlett:2001jj,
    author = "Gauntlett, Jerome P. and Kim, Nakwoo",
    title = "{M five-branes wrapped on supersymmetric cycles 2}",
    eprint = "hep-th/0109039",
    archivePrefix = "arXiv",
    reportNumber = "QMUL-PH-01-08",
    doi = "10.1103/PhysRevD.65.086003",
    journal = "Phys. Rev. D",
    volume = "65",
    pages = "086003",
    year = "2002"
}

@article{Gauntlett:2002rv,
    author = "Gauntlett, Jerome P. and Kim, Nakwoo and Pakis, Stathis and Waldram, Daniel",
    title = "{M theory solutions with AdS factors}",
    eprint = "hep-th/0202184",
    archivePrefix = "arXiv",
    reportNumber = "QMUL-PH-02-04",
    doi = "10.1088/0264-9381/19/15/305",
    journal = "Class. Quant. Grav.",
    volume = "19",
    pages = "3927--3946",
    year = "2002"
}

@article{Nastase:1999kf,
    author = "Nastase, Horatiu and Vaman, Diana and van Nieuwenhuizen, Peter",
    title = "{Consistency of the AdS(7) x S(4) reduction and the origin of selfduality in odd dimensions}",
    eprint = "hep-th/9911238",
    archivePrefix = "arXiv",
    reportNumber = "ITP-SB-99-56",
    doi = "10.1016/S0550-3213(00)00193-0",
    journal = "Nucl. Phys. B",
    volume = "581",
    pages = "179--239",
    year = "2000"
}

@article{Nastase:1999cb,
    author = "Nastase, Horatiu and Vaman, Diana and van Nieuwenhuizen, Peter",
    title = "{Consistent nonlinear K K reduction of 11-d supergravity on AdS(7) x S(4) and selfduality in odd dimensions}",
    eprint = "hep-th/9905075",
    archivePrefix = "arXiv",
    reportNumber = "ITP-SB-99-19",
    doi = "10.1016/S0370-2693(99)01266-6",
    journal = "Phys. Lett. B",
    volume = "469",
    pages = "96--102",
    year = "1999"
}

@article{cvluposatr,
    author = "Cvetic, Mirjam and Lu, Hong and Pope, C. N. and Sadrzadeh, A. and Tran, Tuan A.",
    title = "{S$^3$ and S$^4$ reductions of type IIA supergravity}",
    eprint = "hep-th/0005137",
    archivePrefix = "arXiv",
    reportNumber = "CTP-TAMU-13-00, UPR-888-T",
    doi = "10.1016/S0550-3213(00)00466-1",
    journal = "Nucl. Phys. B",
    volume = "590",
    pages = "233--251",
    year = "2000"
}

@book{notspinc,
    author = "Lawson, H. B. and Michelsohn, M. L.",
    title = "{Spin geometry}",
    publisher = {Princeton University Press},
    SERIES = {Princeton Mathematical Series},
    VOLUME = {38},
    year = "1998"
}

@article{Pernici:1984xx,
    author = "Pernici, M. and Pilch, K. and van Nieuwenhuizen, P.",
    title = "{Gauged maximally extended supergravity in seven-dimensions}",
    reportNumber = "ITP-SB-84-31",
    doi = "10.1016/0370-2693(84)90813-X",
    journal = "Phys. Lett. B",
    volume = "143",
    pages = "103--107",
    year = "1984"
}

@article{Wu,
    author = {Gibson, Cameron and G{\"u}nel, Okan and Larios, Gabriel and Pope, C. N.},
    title = "{Spin$^{h}$ structure, scalar and charged spinor eigenfunctions on the SU(3)/SO(3) Wu manifold}",
    eprint = "2512.19497",
    archivePrefix = "arXiv",
    primaryClass = "hep-th",
    doi = "10.1007/JHEP05(2026)173",
    journal = "JHEP",
    volume = "05",
    pages = "173",
    year = "2026"
}

@article{hawpop,
    author = "Hawking, S. W. and Pope, C. N.",
    title = "{Generalized spin structures in quantum gravity}",
    reportNumber = "Print-78-0374 (CAMBRIDGE)",
    doi = "10.1016/0370-2693(78)90167-3",
    journal = "Phys. Lett. B",
    volume = "73",
    pages = "42--44",
    year = "1978"
}

@article{Birmingham,
    author = "Birmingham, Danny",
    title = "{Topological black holes in Anti-de Sitter space}",
    eprint = "hep-th/9808032",
    archivePrefix = "arXiv",
    doi = "10.1088/0264-9381/16/4/009",
    journal = "Class. Quant. Grav.",
    volume = "16",
    pages = "1197--1205",
    year = "1999"
}

@article{bobev2,
    author = "Bobev, Nikolay and David, Marina and Dimitrov, Vasil and Hong, Junho",
    title = "{AdS$_{7}$ black holes and holography}",
    eprint = "2509.02833",
    archivePrefix = "arXiv",
    primaryClass = "hep-th",
    doi = "10.1007/JHEP11(2025)159",
    journal = "JHEP",
    volume = "11",
    pages = "159",
    year = "2025"
}

@article{bobev1,
    author = "Bobev, Nikolay and David, Marina and Hong, Junho and Mouland, Rishi",
    title = "{AdS$_{7}$ black holes from rotating M5-branes}",
    eprint = "2307.06364",
    archivePrefix = "arXiv",
    primaryClass = "hep-th",
    doi = "10.1007/JHEP09(2023)143",
    journal = "JHEP",
    volume = "09",
    pages = "143",
    year = "2023",
    note = "[Erratum: JHEP 09, 198 (2023)]"
}

@article{malda,
    author = "Maldacena, Juan Martin",
    title = "{The large $N$ limit of superconformal field theories and supergravity}",
    eprint = "hep-th/9711200",
    archivePrefix = "arXiv",
    reportNumber = "HUTP-97-A097, HUTP-98-A097",
    doi = "10.4310/ATMP.1998.v2.n2.a1",
    journal = "Adv. Theor. Math. Phys.",
    volume = "2",
    pages = "231--252",
    year = "1998"
}

@article{gubklepol,
    author = "Gubser, S. S. and Klebanov, Igor R. and Polyakov, Alexander M.",
    title = "{Gauge theory correlators from noncritical string theory}",
    eprint = "hep-th/9802109",
    archivePrefix = "arXiv",
    reportNumber = "PUPT-1767",
    doi = "10.1016/S0370-2693(98)00377-3",
    journal = "Phys. Lett. B",
    volume = "428",
    pages = "105--114",
    year = "1998"
}

@article{withol,
    author = "Witten, Edward",
    title = "{Anti de Sitter space and holography}",
    eprint = "hep-th/9802150",
    archivePrefix = "arXiv",
    reportNumber = "IASSNS-HEP-98-15",
    doi = "10.4310/ATMP.1998.v2.n2.a2",
    journal = "Adv. Theor. Math. Phys.",
    volume = "2",
    pages = "253--291",
    year = "1998"
}

@article{holreview,
    author = "Bobev, Nikolay and Crichigno, P. Marcos",
    title = "{Universal RG flows across dimensions and holography}",
    eprint = "1708.05052",
    archivePrefix = "arXiv",
    primaryClass = "hep-th",
    doi = "10.1007/JHEP12(2017)065",
    journal = "JHEP",
    volume = "12",
    pages = "065",
    year = "2017"
}

@article{maldacenanunez,
    author = "Maldacena, Juan Martin and Nunez, Carlos",
    editor = "Duff, Michael J. and Liu, J. T. and Lu, J.",
    title = "{Supergravity description of field theories on curved manifolds and a no go theorem}",
    eprint = "hep-th/0007018",
    archivePrefix = "arXiv",
    doi = "10.1142/S0217751X01003937",
    journal = "Int. J. Mod. Phys. A",
    volume = "16",
    pages = "822--855",
    year = "2001"
}

@article{zayas,
    author = "Hong, Junho and Macpherson, Niall T. and Pando Zayas, Leopoldo A.",
    title = "{Aspects of AdS$_{2}$ classification in M-theory: solutions with mesonic and baryonic charges}",
    eprint = "1908.08518",
    archivePrefix = "arXiv",
    primaryClass = "hep-th",
    reportNumber = "LCTP-19-18",
    doi = "10.1007/JHEP11(2019)127",
    journal = "JHEP",
    volume = "11",
    pages = "127",
    year = "2019"
}

@article{bubbling,
    author = "Mac Conamhna, Oisin A. P. and O Colgain, Eoin",
    title = "{Supersymmetric wrapped membranes, AdS$_2$ spaces, and bubbling geometries}",
    eprint = "hep-th/0612196",
    archivePrefix = "arXiv",
    reportNumber = "IMPERIAL-TP-2006-OC-02",
    doi = "10.1088/1126-6708/2007/03/115",
    journal = "JHEP",
    volume = "03",
    pages = "115",
    year = "2007"
}

@article{cosets,
    author = "Gibbons, G. W. and L{\"u}, H. and Pope, C. N.",
    title = {Einstein Metrics on Group Manifolds and Cosets},
    eprint = "0903.2493",
    archivePrefix = "arXiv",
    primaryClass = "hep-th",
    reportNumber = "DAMTP-2009-20, MIFP-09-09",
    doi = "10.1016/j.geomphys.2011.01.004",
    journal = "J. Geom. Phys.",
    volume = "61",
    pages = "947--960",
    year = "2011"
}

@article{gonzalez,
  title={Invariant contact metric structures on tangent sphere bundles of compact symmetric spaces},
  author={Gonz{\'a}lez-D{\'a}vila, JC},
  journal={Revista de la Real Academia de Ciencias Exactas, F{\'\i}sicas y Naturales. Serie A. Matem{\'a}ticas},
  volume={117},
  number={4},
  pages={142},
  year={2023},
  publisher={Springer},
  doi = "10.1007/s13398-023-01475-x"
}

@article{embeddingkiller,
    author = "Bonetti, Federico and Minasian, Ruben and Camell, Valent{\'\i} Vall and Weck, Peter",
    title = "{Consistent truncations from the geometry of sphere bundles}",
    eprint = "2212.08068",
    archivePrefix = "arXiv",
    primaryClass = "hep-th",
    reportNumber = "LMU-ASC 54/22",
    doi = "10.1007/JHEP05(2023)156",
    journal = "JHEP",
    volume = "05",
    pages = "156",
    year = "2023"
}

@book{MilnorStasheff1974,
      author    = {Milnor, John W. and Stasheff, James D.},
      title     = {Characteristic Classes},
      series    = {Annals of Mathematics Studies},
      number    = {76},
      publisher = {Princeton University Press},
      address   = {Princeton, NJ},
      year      = {1974},
    }

@book{BottTu1982,
      author    = {Bott, Raoul and Tu, Loring W.},
      title     = {Differential Forms in Algebraic Topology},
      series    = {Graduate Texts in Mathematics},
      number    = {82},
      publisher = {Springer-Verlag},
      address   = {New York},
      year      = {1982},
      doi       = {10.1007/978-1-4757-3951-0},
    }

@book{Hatcher2002,
      author    = {Hatcher, Allen},
      title     = {Algebraic Topology},
      publisher = {Cambridge University Press},
      address   = {Cambridge},
      year      = {2002},
      isbn      = {978-0-521-79540-1},
      note      = {Available at \url{https://pi.math.cornell.edu/~hatcher/AT/AT.pdf}},
    }

@book{SteenrodEpstein1962,
      author    = {Steenrod, Norman E. and Epstein, David B. A.},
      title     = {Cohomology Operations},
      series    = {Annals of Mathematics Studies},
      number    = {50},
      publisher = {Princeton University Press},
      address   = {Princeton, NJ},
      year      = {1962},
    }

@article{debray2,
    author = "Debray, Arun and Yu, Matthew",
    title = "{What bordism-theoretic anomaly cancellation can do for U}",
    eprint = "2210.04911",
    archivePrefix = "arXiv",
    primaryClass = "hep-th",
    doi = "10.1007/s00220-024-04937-4",
    journal = "Commun. Math. Phys.",
    volume = "405",
    number = "7",
    pages = "154",
    year = "2024"
}

@article{Bar65,
    AUTHOR = {Barden, D.},
     TITLE = {Simply connected five-manifolds},
   JOURNAL = {Ann. of Math. (2)},
  FJOURNAL = {Annals of Mathematics. Second Series},
    VOLUME = {82},
      YEAR = {1965},
     PAGES = {365--385},
      ISSN = {0003-486X},
   MRCLASS = {57.10},
  MRNUMBER = {184241},
MRREVIEWER = {S. Smale},
       DOI = {10.2307/1970702},
       URL = {https://doi.org/10.2307/1970702},
}

@article{Flo73,
    AUTHOR = {Floyd, E. E.},
     TITLE = {The number of cells in a non-bounding manifold},
   JOURNAL = {Ann. of Math. (2)},
  FJOURNAL = {Annals of Mathematics. Second Series},
    VOLUME = {98},
      YEAR = {1973},
     PAGES = {210--225},
      ISSN = {0003-486X},
   MRCLASS = {57D90},
  MRNUMBER = {334256},
MRREVIEWER = {J. W. Vick},
       DOI = {10.2307/1970782},
}

@article{Lee:2020ojw,
    author = "Lee, Yasunori and Ohmori, Kantaro and Tachikawa, Yuji",
    title = "{Revisiting Wess-Zumino-Witten terms}",
    eprint = "2009.00033",
    archivePrefix = "arXiv",
    primaryClass = "hep-th",
    reportNumber = "IPMU-20-0093",
    doi = "10.21468/SciPostPhys.10.3.061",
    journal = "SciPost Phys.",
    volume = "10",
    number = "3",
    pages = "061",
    year = "2021"
}

@article{split,
  title="{Metrics of positive Ricci curvature on simply-connected manifolds of dimension 6k}",
  author={Reiser, Philipp},
  journal={Journal of Topology},
  volume={17},
  number={4},
  pages={e70007},
  year={2024},
  publisher={Wiley Online Library},
  doi={10.1112/topo.70007}
}

@article{reiser,
  title="{Generalized surgery on Riemannian manifolds of positive Ricci curvature}",
  author={Reiser, Philipp},
  journal={Trans. Amer. Math. Soc},
  volume={376},
  number={5},
  pages={3397--3418},
  year={2023},
  doi={10.1090/tran/8789}
}

@article{class1,
    author = "Kim, Nakwoo and Park, Jong-Dae",
    title = "{Comments on AdS$_2$ solutions of D=11 supergravity}",
    eprint = "hep-th/0607093",
    archivePrefix = "arXiv",
    doi = "10.1088/1126-6708/2006/09/041",
    journal = "JHEP",
    volume = "09",
    pages = "041",
    year = "2006"
}

@article{class2,
    author = "Gauntlett, Jerome P. and Kim, Nakwoo and Waldram, Daniel",
    title = "{Supersymmetric AdS$_3$, AdS$_2$ and bubble solutions}",
    eprint = "hep-th/0612253",
    archivePrefix = "arXiv",
    reportNumber = "IMPERIAL-TP-2006-JG-03",
    doi = "10.1088/1126-6708/2007/04/005",
    journal = "JHEP",
    volume = "04",
    pages = "005",
    year = "2007"
}

@article{class3,
    author = "Donos, Aristomenis and Gauntlett, Jerome P. and Kim, Nakwoo",
    title = "{AdS solutions through transgression}",
    eprint = "0807.4375",
    archivePrefix = "arXiv",
    primaryClass = "hep-th",
    reportNumber = "DESY-08-104, IMPERIAL-TP-2008-JG-01",
    doi = "10.1088/1126-6708/2008/09/021",
    journal = "JHEP",
    volume = "09",
    pages = "021",
    year = "2008"
}

@article{class4,
    author = "Donos, Aristomenis and Gauntlett, Jerome P.",
    title = "{Supersymmetric quantum criticality supported by baryonic charges}",
    eprint = "1208.1494",
    archivePrefix = "arXiv",
    primaryClass = "hep-th",
    reportNumber = "IMPERIAL-TP-2012-JG-04",
    doi = "10.1007/JHEP10(2012)120",
    journal = "JHEP",
    volume = "10",
    pages = "120",
    year = "2012"
}

@article{11dspinor1,
    author = "Gauntlett, Jerome P. and Pakis, Stathis",
    title = "{The geometry of D = 11 Killing spinors}",
    eprint = "hep-th/0212008",
    archivePrefix = "arXiv",
    reportNumber = "QMUL-PH-02-22",
    doi = "10.1088/1126-6708/2003/04/039",
    journal = "JHEP",
    volume = "04",
    pages = "039",
    year = "2003"
}

@article{11dspinor2,
    author = "Gauntlett, Jerome P. and Gutowski, Jan B. and Pakis, Stathis",
    title = "{The geometry of D = 11 null Killing spinors}",
    eprint = "hep-th/0311112",
    archivePrefix = "arXiv",
    reportNumber = "IMPERIAL-TP-03-04-3",
    doi = "10.1088/1126-6708/2003/12/049",
    journal = "JHEP",
    volume = "12",
    pages = "049",
    year = "2003"
}

@article{Katmadas:2015ima,
    author = "Katmadas, Stefanos and Tomasiello, Alessandro",
    title = "{AdS$_{4}$ black holes from M-theory}",
    eprint = "1509.00474",
    archivePrefix = "arXiv",
    primaryClass = "hep-th",
    doi = "10.1007/JHEP12(2015)111",
    journal = "JHEP",
    volume = "12",
    pages = "111",
    year = "2015"
}

@article{Aloff,
  title="{An infinite family of distinct 7-manifolds admitting positively curved Riemannian structures}",
  author={Simon Aloff and Nolan Wallach},
  journal={Bulletin of the American Mathematical Society},
  year={1975},
  volume={81},
  pages={93-97},
  doi={https://doi.org/10.1090/S0002-9904-1975-13649-4}
}

@article{fof,
    author = "Figueroa-O'Farrill, Jose",
    title = "{Symmetric M-theory backgrounds}",
    eprint = "1112.4967",
    archivePrefix = "arXiv",
    primaryClass = "hep-th",
    reportNumber = "EMPG-11-28",
    doi = "10.2478/s11534-012-0160-6",
    journal = "Central Eur. J. Phys.",
    volume = "11",
    pages = "1--36",
    year = "2013"
}

@article{lu2023nonlinear,
    author = "Lu, Da-Chuan",
    title = "{Nonlinear sigma model description of deconfined quantum criticality in arbitrary dimensions}",
    eprint = "2209.00670",
    archivePrefix = "arXiv",
    primaryClass = "cond-mat.str-el",
    doi = "10.21468/SciPostPhysCore.6.3.047",
    journal = "SciPost Phys. Core",
    volume = "6",
    pages = "047",
    year = "2023"
}

@article{Douglas:2010iu,
    author = "Douglas, Michael R.",
    title = "{On D=5 super Yang-Mills theory and (2,0) theory}",
    eprint = "1012.2880",
    archivePrefix = "arXiv",
    primaryClass = "hep-th",
    doi = "10.1007/JHEP02(2011)011",
    journal = "JHEP",
    volume = "02",
    pages = "011",
    year = "2011"
}

@article{Klebanov:1996un,
    author = "Klebanov, Igor R. and Tseytlin, Arkady A.",
    title = "{Entropy of near extremal black p-branes}",
    eprint = "hep-th/9604089",
    archivePrefix = "arXiv",
    reportNumber = "PUPT-1613, IMPERIAL-TP-95-96-39",
    doi = "10.1016/0550-3213(96)00295-7",
    journal = "Nucl. Phys. B",
    volume = "475",
    pages = "164--178",
    year = "1996"
}

@article{Henningson:1998gx,
    author = "Henningson, M. and Skenderis, K.",
    title = "{The holographic Weyl anomaly}",
    eprint = "hep-th/9806087",
    archivePrefix = "arXiv",
    reportNumber = "CERN-TH-98-188, KUL-TF-98-21",
    doi = "10.1088/1126-6708/1998/07/023",
    journal = "JHEP",
    volume = "07",
    pages = "023",
    year = "1998"
}

@article{crowley11,
  title={5-manifolds: 1-connected},
  author={Crowley, Diarmuid},
  journal={Bulletin of the Manifold Atlas},
  pages={49--55},
  year={2011}
}






\end{document}